\pdfoutput=1
\documentclass[a4paper,fleqn,usenatbib]{mnras}

\usepackage[T1]{fontenc}
\usepackage{ae,aecompl}

\DeclareRobustCommand{\VAN}[3]{#2}
\let\VANthebibliography\thebibliography
\def\thebibliography{\DeclareRobustCommand{\VAN}[3]{##3}\VANthebibliography}
\usepackage{graphicx}	% Including figure files
\usepackage{amsmath}	% Advanced maths commands
\usepackage{amssymb}	% Extra maths symbols
\usepackage{verbatim}

\newcommand{\g}{${\gamma = -2.11 \pm 0.03}$}
\newcommand{\s}{${\gamma_{\star} = -2.86 \pm 0.05}$}
\newcommand{\re}{$\text{R}_{\text{e}}$}

\usepackage{newtxtext}
\usepackage[varvw]{newtxmath}

\title[Density slopes of early-type galaxies]{Total mass density slopes of early-type galaxies using Jeans dynamical modelling at redshifts $0.29 < z< 0.55$}
 
\author[Derkenne et al.]{
Caro Derkenne,$^{1,2}$\thanks{E-mail: caro.derkenne@hdr.mq.edu.au}
Richard M. McDermid,$^{1,2}$
Adriano Poci,$^{3,1}$
Rhea-Silvia Remus,$^{4}$ \newauthor
Inger J\o rgensen,$^{5}$
and Eric Emsellem$^{6,7}$
\\
$^{1}$ Research Centre for Astronomy, Astrophysics, and Astrophotonics, Department of Physics and Astronomy, Macquarie University,
NSW 2109, Australia\\
$^{2}$ ARC Centre of Excellence for All Sky Astrophysics in 3 Dimensions (ASTRO 3D), Australia \\
$^{3}$ Centre for Extragalactic Astronomy, University of Durham, Stockton Road, Durham DH1 3LE, United Kingdom\\
$^{4}$ Universit\"{a}ts-Sternwarte M\"{u}nchen, Fakult\"{a}t f\"{u}r Physik, LMU M\"{u}nchen, Scheinerstr. 1, D-81679 M\"{u}nchen, Germany\\ 
$^{5}$ NSF’s NOIRLab, 670 N. A’ohoku Place, Hilo, Hawai’i, 96720, USA\\
$^{6}$ European Southern Observatory, Karl-Schwarzschild-Str. 2, 85748 Garching, Germany \\
$^{7}$ Univ Lyon, Univ. Lyon 1, ENS de Lyon, CNRS, Centre de Recherche Astrophysique de Lyon UMR5574, 69230 Saint-Genis-Laval France \\ 
}
\date{Accepted XXX. Received YYY; in original form ZZZ}

\pubyear{2020}

\begin{document}
\label{firstpage}
\pagerange{\pageref{firstpage}--\pageref{lastpage}}
\maketitle

\begin{abstract}
  The change of the total mass density slope, $\gamma$, of early-type galaxies through cosmic time is a probe of evolutionary pathways. Hydrodynamical cosmological simulations show that at high redshifts density profiles of early-type galaxies were on average steep ($\gamma \sim -3$). As redshift approaches zero, gas-poor  mergers progressively cause the total mass density slope to approach the `isothermal' slope of $\gamma \sim -2$. Simulations therefore predict steep density slopes at high redshifts, with little to no evolution in density slopes below $z \sim 1$. Gravitational lensing results in the same redshift range find the opposite, namely a significant trend of shallow density slopes at high redshifts, becoming steeper as redshift approaches zero. Gravitational lensing results indicate a different evolutionary mechanism for early-type galaxies than dry merging, such as continued gas accretion or off-axis mergers. At redshift zero, isothermal solutions are obtained by both simulations and dynamical modelling. This work applies the Jeans dynamical modelling technique to observations of galaxies at intermediate redshifts ($0.29 < z < 0.55 $) in order to derive density slopes to address the tension between observations and simulations. We combine two-dimensional kinematic fields from MUSE data with Hubble Space Telescope photometry. The density slopes of 90 early-type galaxies from the Frontier Fields project are presented. The total sample has a median of \g \,\,(standard error), in agreement with dynamical modelling studies at redshift zero. The lack of evolution in total density slopes in the past 4-6 Gyrs supports a dry merging model for early-type galaxy evolution. 
 \end{abstract}


%%%%%%%%%%%%%%%%%%%%%%%%%%%%%%%%%%%%%%%%%%%%%%%%%%

%%%%%%%%%%%%%%%%% BODY OF PAPER %%%%%%%%%%%%%%%%%%

\section{Introduction}
Studies of early-type galaxies are particularly informative for testing predictions of galaxy evolution models. Early-type galaxies are composed of old, passive stellar populations, representing the end product of many of the key processes of galaxy formation and evolution, such as star formation quenching, mergers and accretion events \citep{conselice_evolution_2014}. More pragmatically, they are also massive, have relatively high surface brightness, and are generally free of dust, allowing for easy observations even at high redshifts.

The mass assembly of early-type galaxies across cosmic time is a combination of the accretion of stars, and the accretion of gas. Each process leaves a fingerprint on the structure and dynamics of the galaxy, by which its evolutionary pathway can be traced. In the case of stellar, or `dry' accretion, stars that have formed external to the main progenitor galaxy are accreted. The increase in mass is accompanied by a significant increase in galaxy radius, implying a reduction in mean density, or ``puffing-up'' of the galaxy \citep{naab_minor_2009, bezanson_relation_2009, van_dokkum_growth_2010,hilz_relaxation_2012,oogi_dry_2013}.

In contrast, the accretion of gas rich satellite galaxies, or a `wet' merger, may lead to only a minor increase in galaxy radius due to the dissipative nature of the gas \citep{barnes_1991_fueling,mihos_ultraluminous_1994}. Depending on the detailed merger configuration (and gas richness), part of the gas may contribute to a relatively compact central starburst, increasing the mass density in the inner kpc \citep{sanders_1996_luminous}, or more distributed star formation \citep{whitmore_1995_hubble,renaud_2008_starbust}.

Cosmological simulations indicate a two-phase evolution process for early-type galaxies \citep{white_core_1978,blumenthal_formation_1984,oser_two_2010}.
In-situ star formation within the host galaxy dominates at high redshifts, followed by a period of mostly dissipationless mergers. Dry mergers promote inside-out growth, as stellar material accreted at large radii leaves the central regions of the galaxy unaffected \citep{bezanson_spatially_2018,hilz_how_2013,karademir_outer_2019}. In terms of the mass distribution of a galaxy, the two-phase model of late dry merger events would indicate that as redshift approaches zero, galaxies tend towards shallower total density gradients, with the density expressed by {$\rho(r) \propto r^{\gamma}$}.

This two-phase evolution is supported by simulation studies which have traced the total mass density slopes through cosmic time. Total mass density slopes $\gamma$ are found to be steeper at greater lookback times, and approximately constant below redshifts of $z \sim 1$, calculated over $\sim 4$ effective radii ($\text{R}_\text{e}$) \citep{remus_dark_2013,remus_co-evolution_2017,xu_inner_2017,springel_first_2018,wang_early-type_2019-1}. At redshifts of zero, the simulations indicate a convergence of density slopes around $\gamma \approx -2$, possibly as this `isothermal' solution represents a low energy, high entropy state that galaxies evolve towards due to multiple stellar accretion events. 

Observations of the total mass density slopes in the local Universe agree well with the predictions of simulations. Density slopes at low redshifts can be probed via multiple methods, such as tracing HI gas, the dynamics of planetary nebulae, and analysing the temperature of X-Ray gas \citep{coccato_kinematic_2009, weijmans_shape_2008, humphrey_slope_2010,serra_linear_2016}. The advent of integral field spectroscopy has provided access to detailed spatially-resolved maps of stellar kinematics, allowing the application of more general dynamical modelling approaches that are better suited to the shapes and orbital anisotropies of real galaxies.

A small sample of local ($< 30$ Mpc), fast rotating early-type galaxies from the SLUGSS survey \citep{brodie_sages_2014} was used to derive total density slopes from Jeans modelling, finding ${\overline{\gamma} = - 2.19 \pm 0.03}$ with a small intrinsic scatter of ${\sigma_{\gamma} = 0.11}$ \citep{cappellari_small_2015}.
\citet{poci_systematic_2017} used a volume-limited sample of 258 early-type galaxies from the ${\text{ATLAS}^{\text{3D}}}$ survey \citep{cappellari_atlas3d_2011}, reporting a mean of ${\overline{\gamma} = -  2.193 \pm 0.016}$ for galaxies with a velocity dispersion above ${\sim 130 \mathrm{\,\,kms}^{-1}}$, with an intrinsic scatter of ${\sigma_{\gamma} = 0.168}$. This sample extended to $\sim 1\text{R}_\text{e}$. $\text{ATLAS}^{\text{3D}}$ was combined with SLUGSS data by \citet{bellstedt_sluggs_2018}, with an obtained density slope mean of $\overline{\gamma} = -2.12 \pm 0.05$, for an overlapping sample with \citet{poci_systematic_2017} but with a greater radial range ($\sim 4\text{R}_\text{e}$). A sample of 2778  early-type and spiral galaxies in the local Universe from the MaNGA survey \citep{bundy_overview_2014} found the highest mean density slope in the local Universe, with ${\overline{\gamma} = -2.24}$, with an intrinsic scatter of $\sigma_{\gamma} = 0.22$ \citep{li_sdss-iv_2019}. However, this sample was not restricted to early-types. At a redshift of $\sim 0$, there is consensus on slightly steeper than isothermal density profiles. 

At intermediate redshifts, density slopes are observationally probed by gravitational lensing, which provides a total integrated mass measurement within the Einstein radius based on how a more distant object's light is bent around the lens \citep{meylan_gravitational_2006}.  Studies of lensing systems in the Lenses Structures and Dynamics (LSD) survey \citep{koopmans_lenses_2004-1} found shallower ($\gamma \sim -1.75$) than isothermal mean slopes, from combining lensing mass measurements at large radii with mass measurements at small radii inferred from slit spectroscopy aperture velocity dispersions to model the total potential \citep{treu_massive_2004,koopmans_sloan_2006}. The slopes were for lensing systems across the redshift range $ 0.5 < z < 1$. A similar approach with a larger sample was conducted by \citet{auger_sloan_2010} using the Sloan Lens ACS Survey (SLACS), finding ${\overline{\gamma} =  - 2.08 \pm 0.03}$ for the redshift range ${0 < z < 0.36}$. Orbital isotropy was enforced in the models, which is not reflective of real galaxies \citep{cappellari_sauron_2007}.  

\citet{barnabe_two-dimensional_2011} combined gravitational lensing data with 2D stellar kinematic maps and two-integral Schwarzschild modelling to characterise a total mass potential, optimising density profiles with a bayesian inference approach. In these models, orbital anisotropy was also allowed. The obtained mean density slope from the 16 lens systems in this study is $\overline{\gamma} = - 2.07 \pm 0.04$, for redshifts around $z \sim 0.2$, similar to \citet{shajib_dark_2021}. Subsequent studies using combinations of data from SLACS, the LSD survey, the Strong Lenses in the Legacy Survey \citep[S2LS;][]{ruff_sl2s_2011}, and the BOSS Emission-Line Lens Survey \citep{brownstein_boss_2011}, found mild to significant trends with redshift, in that total density slopes were found to be systematically steeper at lower redshifts \citep{ruff_sl2s_2011,bolton_boss_2012,sonnenfeld_sl2s_2013}. Contrary to the predictions of simulations, \citet{bolton_boss_2012} found the most significant trend of shallow slopes at high redshifts to steeper slopes in the local Universe, of $d\langle\gamma\rangle/dz = 0.6 \pm 0.15$ for the redshift range $0 < z< 0.7$. Steepening slopes as redshift approaches zero are suggestive of the continued importance of dissipative processes in galaxy evolution \citep{sonnenfeld_purely_2014}, or potentially the occurrence of off-axis major mergers \citep{bolton_boss_2012}.

The evolution with redshift implied by gravitational lensing results is at odds with the predictions of simulations, which indicate steeper slopes with higher redshifts \citep{remus_dark_2013,remus_co-evolution_2017,wang_early-type_2019-1}, and is also inconsistent to the observed mass-size growth of early-type galaxies with redshift as inferred from stellar light \citep{franx_structure_2008,van_dokkum_growth_2010,van_der_wel_3d-hstcandels_2014,mowla_cosmos-dash_2019}. Furthermore, gravitational lensing is necessarily biased towards the most massive, dense objects which act as effective lenses, and this can impact knowledge of the distribution of total mass density slopes \citep{mandelbaum_galaxy_2009}.
 
This works presents the results of Jeans anisotropic modelling on a sample of intermediate redshift galaxies from the Frontier Fields clusters, in order to determine if there exists a change in the distributions of density slopes at higher redshifts using identical methods to those of local Universe studies. Data from the Multi Unit Spectroscopic Explorer (MUSE) is used in combination with Hubble Space Telescope (HST) photometry to construct dynamical  models by fitting 2D kinematic maps for 90 galaxies across the redshift range $0.29 < z < 0.55$. Section \ref{datasample} describes the data sample obtained from the archives. Section \ref{methods} outlines the methods by which the kinematic fields and stellar potentials were determined, and  the Jeans model definitions. Results are given in Section \ref{results}, and Section \ref{discussion} places them in context with other studies. Conclusions are given in Section \ref{concs}. Throughout, a standard, flat cosmology is adopted with Wilkinson Microwave Anistropy Probe 7 values of  $H_0 = 70.2 \text{ kms}^{-1}\text{ Mpc}^{-1}$,   $\Omega_{\Lambda} = 0.728$, and $\Omega_{m} = 0.272$ \citep{larson_seven-year_2011}.

\section{Data Sample}
\label{datasample}
 The six massive galaxy clusters comprising the Frontier Fields \citep{lotz_frontier_2017} project represent a natural choice for building a sample of early-type galaxies at intermediate redshifts. Each cluster has overlapping Hubble Space Telescope and MUSE fields, and is rich in early-type galaxies suitable for deriving stellar kinematics and subsequent Jeans modelling. The early-type sample aims to be comparable with low redshift samples such as $\text{ATLAS}^{\text{3D}}$ \citep{cappellari_atlas3d_2011} in terms of galaxy masses and sizes. Dense cluster environments have been shown to lead to truncation of the dark matter halos of early-type galaxies in comparison to isolated field galaxies at the same redshift \citep{limousin_truncation_2007,eichner_galaxy_2013}. However, this truncation occurs on scales outside the kinematic range of the data used to constrain the models used in this work, and is therefore not expected to strongly impact our results. We discuss the potential impact of environment further in Section \ref{discussion}.

The Frontier Fields project aims to observe massive galaxy clusters that act as lenses to even more distant objects, providing glimpses of the very early Universe. To that end, over 840 orbits of the Hubble Space Telescope were devoted to image the clusters in seven optical to near-infrared bands, being \textit{F435W}, \textit{F606W}, \textit{F814W}, \textit{F105W}, \textit{F125W}, \textit{F140W}, and \textit{F160W}. In the \textit{F814W} band used in this work, the target field exposure time is $105$ ks, and the drizzled fields result in a spatial sampling of \(0.03\ {\rm arcsec/pixel}\). 

All Hubble Space Telescope processed data was obtained from the STSci MAST Archive.\footnote{10.17909/T9KK5N} For this project, already reduced Epoch 1 or 2, version 1.0 drizzled science images were used. Details of the reduction process for the Frontier Fields HST images are given by \citet{lotz_frontier_2017}, with a brief summary as follows. The data were reduced by masking sky artefacts and aligning to a standard astrometric grid, as well as calibrated with darks, flats, and bias fields. Where possible, fields were stacked to improve image depth. 

The Frontier Fields clusters have over-lapping, albeit smaller, MUSE/VLT fields. For the wide-field mode of MUSE used, the covered wavelength range is 480-930 nm, with a spatial scale of 0.2" per pixel over a 1 $\mathrm{armin}^2$ field. The spectral resolution is $\sim 50 \mathrm{kms}^{-1}$ for $\sim 6590$ \AA \citep{vaughan_stellar_2018}. The sky-region is split into 24 sub-fields via an advanced slicer, which then feed into 24 spectrographs; for details, see \citet{bacon_muse_2010}. Reduced science images from the MUSE-DEEP collection, in the form of 3D data and variance cubes, were obtained from the public ESO Archive Science Portal.\footnote{http://archive.eso.org/scienceportal/home} Briefly, the reduction involves bias removal, flat-fielding, astrometric calibration, flux calibration, and sky-subtraction. For details of the reduction pipeline, see \citet{weilbacher_data_2020}. The data was taken without the aid of adaptive optics, with the point spread function (PSF) full width half maximums (FWHM) given in Table \ref{datasetwhere}, along with integration times and dataset IDs for the ESO Archive.

The Frontier Fields cluster M0717.5+3745 was not used in this work as it has no data in the optical band. The properties  of the remaining five clusters - Abell 2744 (A2744), Abell S1063 (AS1063), Abell 370 (A370), MACS J0416.1-2403 (M0416), and MACS J1149.5+223  (M1149) - are summarised in Table \ref{datasetprops}, including the source used to determine cluster members. 
 
\begin{table*}
  \centering
  \caption{The ESO Science Archive ID for each of the datasets used in this work is provided, along with the MUSE program IDs under which the data was originally obtained. The used PSF FWHM for the MUSE data are given, with further discussion in Section \ref{jean}.}
\label{datasetwhere}
    \begin{tabular}{lcccc} \hline
      Cluster & ESO Data ID & MUSE Program(s) & Effective Exposure Time (s) & PSF FWHM (") \\\hline
      A2744 & ADP.2017-03-24T12:14:09.100 & 095.A-0181, 0.96.A-0496, 0.94.A-0115  & 16345 & 0.64\\ 
      AS1063 & ADP.2017-03-28T12:46:01.331 & 095.A-0653  & 15885 & 1.08 \\ 
      & ADP.2017-03-23T15:58:03.937 & 60.A-9345.& 7877 & 1.45 \\ 
      A370 & ADP.2017-06-06T13:13:38.674 & 096.A-0710, 0.94.A-0115 & 13655 & 0.72  \\
      M0416 &  ADP.2019-10-09T11:36:01.797 & 0.100.A-0763 & 39545 & 0.740 \\ 
      & ADP.2017-03-24T16:19:17.624 & 0.94.A-0525 & 36113 & 0.72  \\ 
      M1149 &ADP.2017-03-24T16:26:09.634 & 294.A-5032 & 15282 & 1.40   \\\hline

	\end{tabular}
\end{table*}

\begin{table*}
\centering
\caption{Data source to determine cluster members for each of the Frontier Fields galaxy clusters used in this work, shown in redshift order. The `Members' column refers to galaxies within the archival MUSE footprint for each cluster, with the members drawn from the given source. The ID numbers used in this work correspond to the IDs in the given source, except for AS1063 and M1149, where the IDs correspond to those in \citet{tortorelli_kormendy_2018}.  For A2744, the mass is a virial estimate within 1.3 Mpc \citep{merten_creation_2011}; For AS1063, the given mass is an M500 estimate \citep{williamson_sunyaev-zeltextquotesingledovich-selected_2011}; For A370, the given mass is a virial estimate \citep{richard_abell_2010}; For M0416, the mass is a total mass within 950 kpc \citep{grillo_clash-vlt_2015-1}; For M1149, the given mass is a total estimate \citep{zheng_magnified_2012}.}
\label{datasetprops}
    \begin{tabular}{lccccc} \hline
      Cluster & Central coordinates (J2000)& Mass ($\times 10 ^{15} \text{ M}_{\odot}$) & Members & Cluster $z$ & Source\\\hline
      A2744 &00:14:21.2, -30:23:50.1 & 1.8 & 156 & 0.308&\citet{mahler_strong_2018}\\
      AS1063 &22:48:44.4, -44:31:48.5&1.2 & 95 & 0.348 & \citet{karman_muse_2015}  \\
      A370 &02:39:52.9, -01:34:36.5&$\sim 1$ & 56 & 0.375& \citet{lagattuta_lens_2017}\\
      M0416 &04:16:08.9, -24:04:28.7&1.4 & 193 & 0.396&  \citet{caminha_refined_2017}\\
      M1149 &11:49:36.3, +22:23:58.1&2.5 & 68 & 0.542&\citet{grillo_story_2016}\\\hline

	\end{tabular}
\end{table*}

\section{Methods}
\label{methods}
\subsection{Stellar kinematics}
\label{kinematics}
The published world coordinates for each galaxy were used to extract individual galaxies from the main MUSE data cube. The size of each individual cube was constructed case-by-case by eye, creating the maximum isolated field possible, on average of extent 10 $\text{R}_\text{e}$, with the minimum case at  $1.8 \text{R}_\text{e}$. To ensure a reasonable signal to noise ratio (SNR) across the field, galaxies were thresholded to a median  SNR of 2 per pixel in the wavelength range $ 5936-6997$ \AA{} for M1149  and $4998-7497$ \AA{} for the other clusters. All galaxies were Voronoi binned \citep{cappellari_adaptive_2003} at a SNR of 10 per spectral pixel per bin for the given wavelength range. We conducted simulations (described in Appendix \ref{appendix:simsect}) of the influence of the number of kinematics bins per field on the measured density slope, and concluded that at least 5 bins per field were necessary. Below this, the density slope could not be reliably constrained. A minimum threshold of at least 5 spatial bins per kinematic field is adopted here. On average the kinematics extend to about 3$\text{R}_\text{e}$ for the targets in the present sample.

The Penalised PiXel-Fitting (pPXF) method \citep{cappellari_parametric_2004,cappellari_improving_2017} was chosen to recover the line of sight velocity distribution (LOSVD). The Medium resolution INT Library of Empirical Spectra (MILES\footnote{http://miles.iac.es/}) single stellar population templates (version 11) \citep{vazdekis_evolutionary_2010} were used to perform the fit, with a standard Salpeter initial mass function slope of 1.3 \citep{salpeter_luminosity_1955}, solar abundance iron content, with spectral full width at half maximum (FWHM) resolution of $2.51$ \AA, and covering the wavelength range $ 3540-7410$ \AA{}.
A central aperture spectrum and optimal template was first created for each galaxy to establish the required subset of templates, as done, for example, by \citet{van_de_sande_sami_2017}, reducing the freedom of the fit (and subsequent scatter in the kinematics) for the lower  SNR bins. All spectra within 1$\text{R}_{\text{e}}$ were co-added to create the central spectrum, using effective radii estimated by using the Source Extraction and Photometry (SEP) library, the Python implementation of Source Extractor \citep{bertin_sextractor:_1996,barbary_sep:_2016}. The central spectrum  median SNR for the final sample, in the fitted wavelength range, is $26$ within the $1\text{R}_{\text{e}}$ aperture, with a maximum in the sample of $120$. 

\begin{figure*}
  \begin{center}
      \includegraphics[width=1.5\columnwidth]{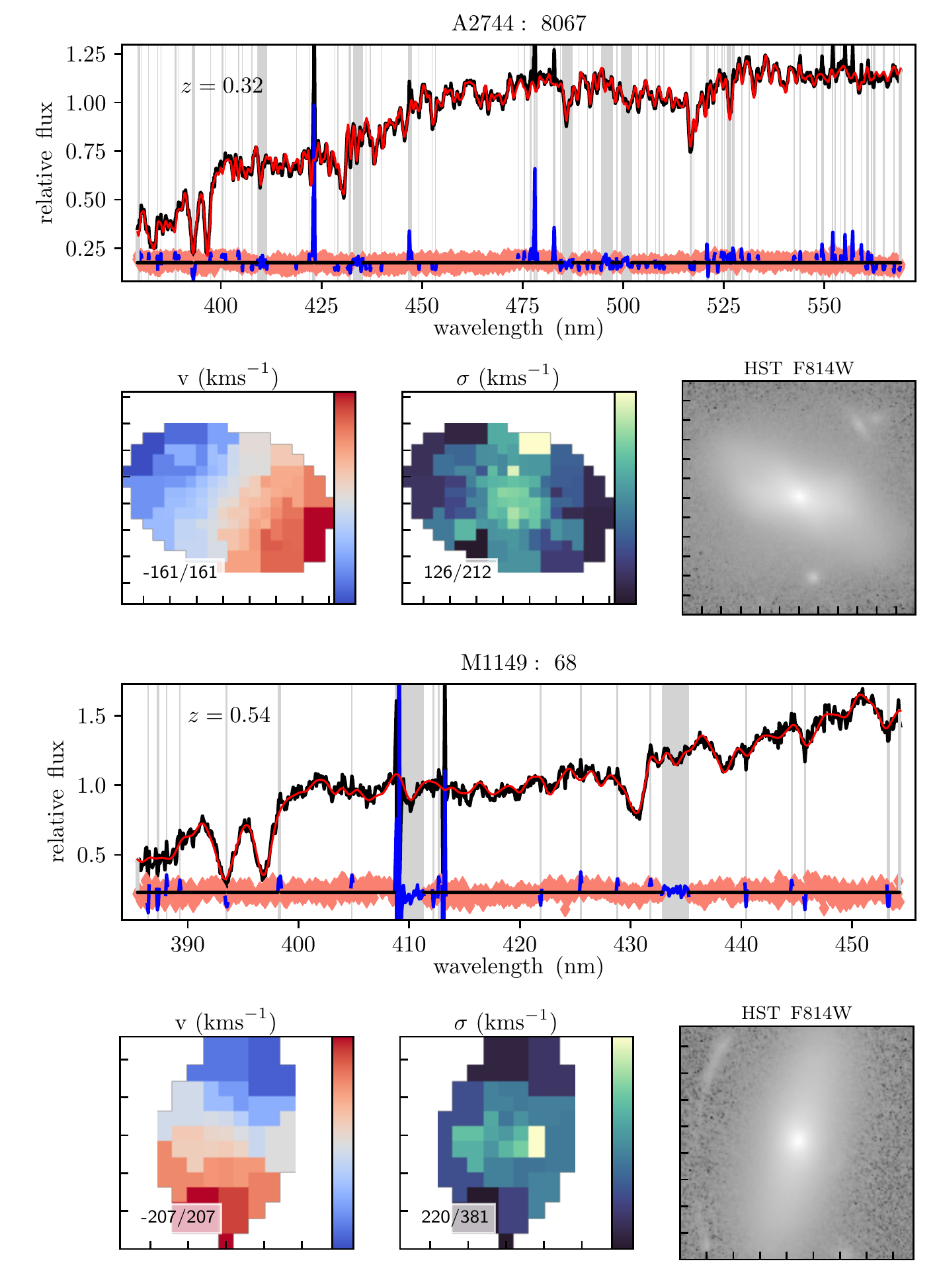}
      \caption{An example of the fields from a low redshift (A2744 8067, $z \sim 0.32$) and a comparatively high redshift (M1149 68, $z \sim 0.54$) galaxy. The rest-frame $1\text{R}_{\text{e}}$ aperture spectrum is shown (black) with the pPXF fit overplotted (red). Residuals (pink) are at the bottom of the panel, with excluded sections of spectrum shown (blue and grey). The velocity and velocity dispersion fields are shown, with the inset numbers indicating the scale of the colours in km/s. The HST \textit{F814W}-band thumbnail of the galaxy is shown. All tick marks correspond to 0.5''. }
      \label{spectrum_figure}
    \end{center}
\end{figure*}

The central aperture spectrum was fitted, in rest frame wavelengths, using pPXF with a 2nd-order Legendre multiplicative polynomial and no additive polynomials. Additive polynomials change absorption line strengths, whilst multiplicative polynomials can correct for spectral calibration issues in the fit. The optimal template was created by the matrix multiplication of the weights of the fit with the input stellar library. No regularisation was performed, as regularisation smooths the weighting of each template's contribution to the optimal template to find star formation histories, which are not needed for dynamical models \citep{cappellari_improving_2017}. The first two moments of the stellar LOSVD were extracted by fitting all Voronoi-binned spectra with this optimal template, with iterative sigma-clipping to produce a `clean' fit. A correction to the velocity dispersion value in each bin was performed after the fit, to account for the change in instrumental resolution of MUSE with wavelength. The correction was made by finding the difference between the broadening at each wavelength as measured for MUSE by \citet{vaughan_stellar_2018}, and the broadening of the MILES library as measured by \citet{falcon-barroso_updated_2011}. This correction term was added in quadrature to the velocity dispersion in each bin.
 Uncertainties for the derived values in each bin were estimated via a Monte Carlo process using 100 trials and shuffling the residuals of the fit for each iteration. All kinematic fields can be seen in Appendix Figure \ref{appendixfigure}. Examples of the obtained kinematic fields, central spectra, and HST images can be seen in Figure \ref{spectrum_figure}. 

\subsection{Photometry}
The photometry analysis presented here is based from HST  data in the \textit{F814W} band. Inferring the intrinsic luminosity distribution within a galaxy from its projected surface brightness involves parametrising the observed light distribution in such a way as to allow different assumed viewing angles to be tested. The multi-Gaussian expansion (MGE) method describes the galaxy surface brightness via a combination of positive, concentric, 2D Gaussians \citep{monnet_modelling_1992,emsellem_multi-gaussian_1994}. Here, we use the python package MgeFit \citet{cappellari_efficient_2002}.\footnote{https://pypi.org/project/mgefit/}
The surface density is described, as in Equation (1) in \citet{cappellari_efficient_2002}, as 
\begin{equation}
  \Sigma(x', y') = \sum^N_{j=1} \frac{L_j}{2\pi \sigma_j^2 q'_j}\exp\Big[-\frac{1}{2\sigma_j^2}\Big(x'^{2}_j + \frac{y'^{2}_j}{q'^{2}_j}\Big)\Big] , 
\end{equation}
where $(x',y')$ are observed (projected) coordinates centred on the galaxy and are aligned with the photometric major axis, $N$ is the number of included Gaussian components, each of luminosity $L_j$; $q'_j$ is axial ratio of the Gaussian component, or flattening; $\sigma_j$ is the dispersion of the Gaussian component measured along the major axis of the galaxy. This parametrisation is conducted in image units - pixels and counts. In this formalism, the position angle ($\psi$) is measured counter-clockwise from the image y-axis (North oriented for this project). The galaxy centre in image coordinates and position angle are both determined by the MgeFit routine. The polar coordinates $R'$ and $\theta '$ are related to the sky coordinates via
\begin{align}
  x'_j &= R'\sin(\theta' - \psi_j) \\
  y'_j & = R'\cos(\theta' - \psi_j).
\end{align}
In the axi-symmetric case, where a galaxy is assumed to be an oblate spheroid, the flattest Gaussian in the MGE model sets the minimum allowed inclination (corresponding to an infinitely thin disk). The roundest Gaussians consistent with the photometric data was fit in order to avoid over-constraining the Jeans dynamical models, as done by \citet{scott_atlas3d_2013-1}. The relation between the Gaussian flattening and inclination is given by
\begin{equation}
  q_j^2 = \frac{q_j^{'2} - \cos^2i}{\sin^2 i}, 
  \label{incs}
\end{equation}
where $q$ is the intrinsic axial ratio and $q'$ is the projected axial ratio on the plane of the sky, with $i$ being the inclination. The deprojected axisymmetric oblate luminosity density is then given by
\begin{equation}
  \nu(R,z)  = \sum^N_{j=1} \frac{L_j}{(\sqrt{2\pi}\sigma_j)^3q_j} \exp\big[-\frac{1}{2\sigma_j^2}\big(R^2 + \frac{z^2}{q_j^2}\big)\big],
  \label{lumindens}
\end{equation}
following Equation (13) of \citet{cappellari_measuring_2008}.
To construct each MGE, individual galaxies were isolated from the main HST field, with masking of adjacent sources performed as necessary. At this step, any galaxy with an irregular or spiral morphology was rejected from the sample. Using the MgeFit package,  the central pixel coordinates, major and minor axes, position angle, and ellipticity were fitted. The galaxy image was divided into sectors spaced linearly in angle and logarithmically in radius; at each sector, the radius, angle, and intensity in counts were recorded. The success of the process  was visually judged by overplotting the fit on the galaxy isophotes in regular magnitude steps, and visually inspecting the corresponding 1D sector fits. The central region of each MGE fit can be seen in Figure \ref{appendixfigure}. The effective radius was derived as the circularised arcsecond extent which contains half the measured luminosity of each MGE, using the total counts from the MGE and the radii in pixel units.  This was converted to physical units (kpc) using a distance estimate to each galaxy based on the mean cluster redshift and the assumed cosmology. This is the effective radius ($\text{R}_{\text{e}}$) reported for each galaxy in Table \ref{full_data} and used in all subsequent analysis. The luminous MGE surface density can remain in arbitrary units as a total potential model setup is used, described in Section \ref{jean}. The luminous MGE dispersion is converted to physical units by multiplying by the HST pixel scale of  $0.03$ arcsec/pixel.

\subsection{Jeans Modelling}
\label{jean}
This project utilises Jeans anisotropic modelling (JAM) through the JamPy package \citep{cappellari_measuring_2008}.\footnote{https://pypi.org/project/jampy/} 
One of the important inclusions of using the JAM technique is accounting for deviation from perfectly isotropic orbital structures, through the anisotropy parameter, $\beta_z$, defined as
\begin{equation}
  \beta_z \equiv 1 - \left(\frac{\sigma_z}{\sigma_R}\right)^2
\end{equation}
where $R$ denotes the radial direction and $z$ is along the axis of symmetry. JamPy models the observed kinematics, predicting the  second moment of the velocity distribution for the luminous barycentres of each projected position - in this case, the Voronoi bin centroids - of the galaxy, integrated along the line of sight. These predictions are derived by deprojecting the MGE model into an intrinsic gravitational potential for a given inclination. The quality of the fit is judged by a chi-square statistic, which takes each data bin as a degree of freedom.
Following \citet{cappellari_measuring_2008}, the JAM formalism makes two key simplifying assumptions: 1) The velocity ellipsoid is aligned with the cylindrical coordinate system ($R$, $z$, $\phi$); and 2) The anistropy is spatially constant, for this implementation of the models. The relationship between the gravitational potential ($\Phi$), luminosity density ($\nu$, defined in Equation \ref{lumindens}), and moments of velocity (\textbf{v}) is then derived from the Jeans equations in the general case as:

\begin{align}
  \frac{\frac{1}{1-\beta_z}\nu\overline{v_z^2} - \nu \overline{v_{\phi}^2}}{R} + \frac{\partial(\frac{1}{1-\beta_z}\nu\overline{v_z^2})}{\partial R} &=  - \nu \frac{\partial \Phi}{\partial R}, \\
  \frac{\partial(\nu\overline{v_z^2})}{\partial z} &= -  \nu \frac{\partial \Phi}{\partial z},
\end{align}
following Equations 8-9 in \citet{cappellari_measuring_2008}. The shorthand notation 
\begin{equation}
  \nu \overline{v_kv_j} \equiv \int v_k v_j f d^3\textbf{v} 
\end{equation}
has been used, with $f$ the distribution function of the stars and the total gravitational potential set in this case by an MGE model, and as defined in \citet{emsellem_multi-gaussian_1994} Equations 39-40. 
%Finally, the dynamical mass-to-light ratio is also a constant, global factor, which scales the entire modelled kinematic fields.
The observed kinematics for each field were combined into a single $v_{\text{rms}}$ field, defined as $v_{\text{rms}} = \sqrt{v^2 + \sigma^2}$. For each individual Voronoi bin there exists a single $v_{\text{rms}}$ value, the sky-plane spatial location of which is set by the luminous barycentre of the bin. Weighting was given to each bin by use of an error vector, defined as 
\begin{equation} 
  \delta v_{\text{rms}} = \frac{\sqrt{(v \times \delta v)^2  + (\sigma \times  \delta \sigma)^2}}{v_{\text{rms}}},
\end{equation}
where the uncertainty on $\mathnormal{v}$ and $\sigma$ are those obtained from the kinematics Monte Carlo process.
The coordinates for each bin were rotated by the galaxy position angle, so that the x-axis aligns with the galaxy projected major axis. Further, the velocity and dispersion fields were bi-symmetrised prior to their combination into a single $v_{\text{rms}}$ field. The symmetrisation reduces outliers that form solutions away from the axisymmetric case, by mapping $x_j \to (x_j, -x_j, x_j, -x_j)$, and similarly for $y_j$. 

The kinematics of each model are convolved with the PSF of the data in order to derive a goodness of fit. A normalised circular MGE model is used to describe the MUSE point spread function (PSF). This was either estimated by fitting stars in the MUSE field, or using the seeing FWHM estimated in the archival data fits header. Both were found to agree within 5\% of each other in cases where stars were available in the field. The effects of the PSF estimation on the density slopes were explored and found to be minimal, for details see Appendix  \ref{appendix:simsect}. 

The gravitational effect of a central super-massive black hole is well below the spatial resolution of the kinematic data. For completeness, however, a central unresolved dark mass component of mass
\begin{multline}
  \log_{10}M_{\text{bh}} = 8.01 + 3.87\log_{10}(\sigma_{\text{e}}/200 \text{ kms}^{-1})  \\ - 0.138 \log_{10}(1 + z)
\end{multline} 
was included in each model, as given in Equation (5) of \citet{robertson_evolution_2006}.
\label{general}

A total generalised potential was implemented for each model, which is representative of early-type galaxy mass distributions, where baryonic matter sits within an extended dark matter halo \citep{white_core_1978,blumenthal_formation_1984}. The general potential used in this work is a double-spherical power law of the form
\begin{equation}
  \rho(r) = \rho_s \left(\frac{r}{r_s}\right)^{\gamma'} \left(\frac{1}{2} + \frac{1}{2}\frac{r}{r_s}\right)^{-\gamma'-3}, 
\end{equation}
where $\gamma'$ and -3 represent the inner and outer total-mass density slopes, separated by the break radius $r_s$, and normalised to $\rho_s$, the density at $r_s$. This is a general form of a so-called `Nuker Law' \citep{lauer_centers_1995}, assuming a Navarro-Frenk-White dark matter halo \citep{navarro_universal_1997}. For this project, as done by \citet{poci_systematic_2017}, the break radius $r_s$ was set at 20kpc, with the maximum radius, $r_m$, set as 25 kpc. This definition ensures that the outer regions can be fixed to a cosmologically-motivated slope without influencing the central regions, which are then free to vary in response to the kinematic data. Arcsecond and parsec scales were converted using the cosmology calculator of \citet{wright_cosmology_2006}.

 A Monte Carlo Markov Chain approach was used for the parameter estimation, via the python package \textsc{emcee} \citep{foreman-mackey_emcee:_2013,hogg_data_2018}.\footnote{https://emcee.readthedocs.io/en/stable/}
The models as implemented here have four free parameters: the inner density slope $\gamma'$, the density at the break radius $\rho_s$, the orbital anisotropy $\beta_z$, and the inclination of the galaxy, $i$. Flat priors were set, with  $-4 < \log_{10} (\rho_s) < 0$, $-3.5 < \gamma' < - 0.5$,  $-0.5 < \beta < 0.5$ and inclination between the minimum angle set by the MGE and $90 ^{\circ}$. These ranges were chosen to be broadly consistent with dynamical modelling conducted at redshift zero \citep{poci_systematic_2017,bellstedt_sluggs_2018}.

A maximum of 10,000 steps was set with 50 independent walkers. A run could terminate earlier if the chain was longer than 50 times the estimated auto-correlation time, providing the auto-correlation time estimates were stable.
\begin{figure}
    \includegraphics[width=\columnwidth]{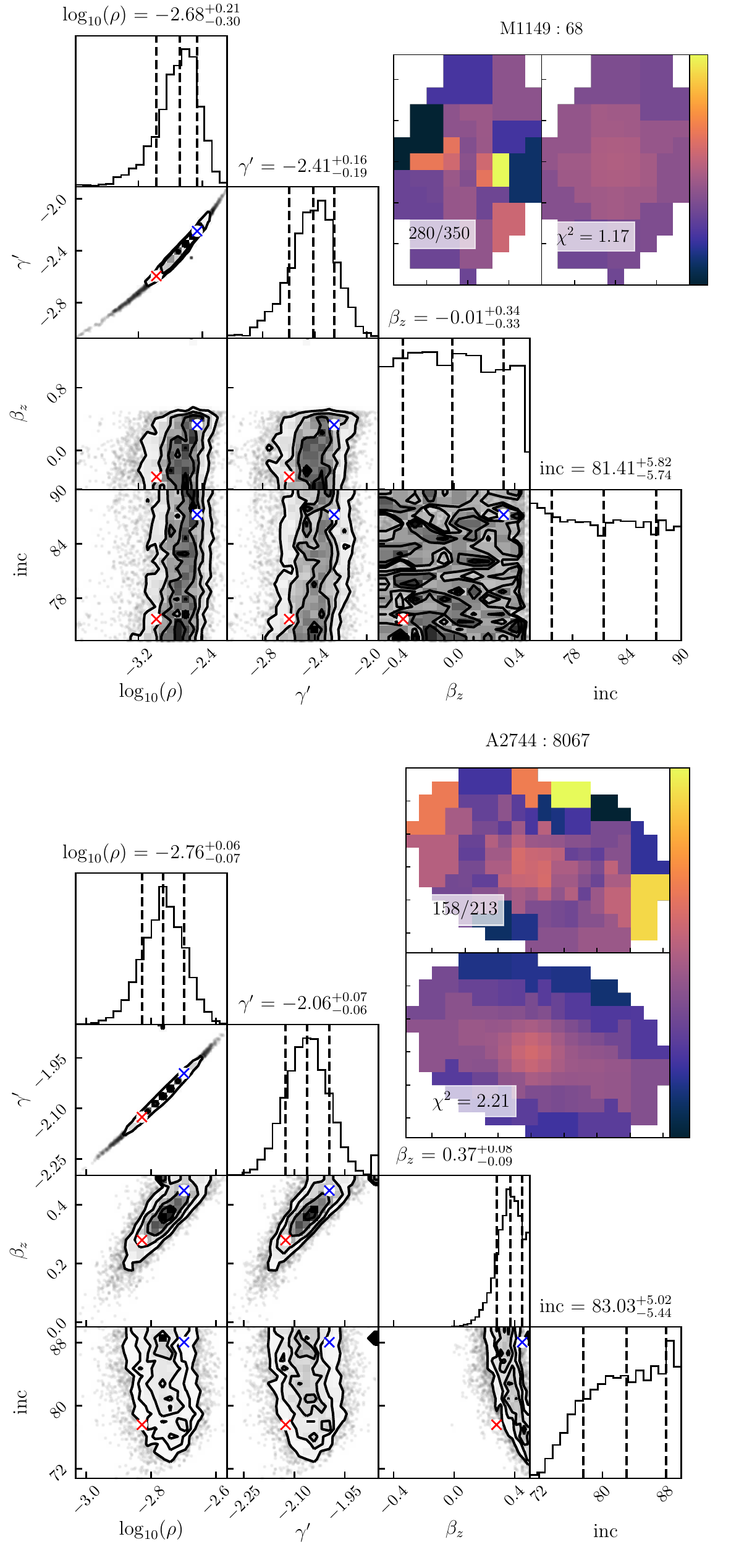}
    \caption{The sampling of the parameter space by the walkers is shown with the marginalised histograms on the top. The observed $v_{\mathrm{rms}}$ field and the modelled field are also shown, for galaxies M1149:68 and A2744:8067 as in Figure \ref{spectrum_figure}. The colour scale is marked on the observed field in units of km/s, and the reduced chi-square value shown on the modelled field. For the two fields, ticks indicate 0.5''. The red and blue crosses are relevant for Figure \ref{appendixfigure}. The red crosses show the parameter values used to construct the 16th percentile $v_{\text{rms}}$ fields (in column 5), and the blue crosses show the parameters used to construct the 84th percentile $v_{\text{rms}}$ fields (in column 7).}
    \label{longcorner}
\end{figure}
The median of the chain for each parameter was used to estimate the most likely parameter value, with the one-sigma uncertainty given by the 16th and 84th percentiles respectively. The best-fit model is defined as the model built from the median values of each parameter chain. Each chain had the initial steps corresponding to twice the galaxy auto-correlation time estimate, or `burn-in', removed. If the median density slope estimate was not more than 2$\sigma$ removed from either boundary, the galaxy was rejected from further analysis, taking the final sample to 90 galaxies across the five clusters. The removal of these poorly constrained cases has not affected the conclusions of this work. A corner plot of the parameter distributions and the resulting median-fit JAM model for two galaxies are shown in Figure \ref{longcorner}. 

As done by \citet{poci_systematic_2017}, the average logarithmic slope of the total mass density profile is defined as 
\begin{equation}
  \label{gammavg}
  \gamma = \frac{\log_{10}\left(\rho(R_{\text{o}})/\rho(R_{\text{i}})\right)}{\log_{10}\left(R_{\text{o}}/R_{\text{i}}\right)},
\end{equation}
with $R_{\text{o}}$ set as the maximum extent of the kinematic data and $R_{\text{i}}$ set as the radius of the MUSE PSF for each cluster.  As also found by \citet{bellstedt_sluggs_2018}, the average slope $\gamma$ is slightly steeper than the \textsc{emcee} optimised slope $\gamma'$. This difference is due to the kinematic data extending on average to 3$\text{R}_{\text{e}}$, where the transition between the inner slope $\gamma'$ and post-break radius slope (set as $-3$) becomes important. 

\section{Results}
\label{results}
The density slopes for the 90 galaxies in the Frontier Fields sample are found using Equation  (\ref{gammavg}). All derived density slopes, aperture dispersions, effective radii and dynamical masses can be found in Table \ref{full_data}. All visual outputs (kinematic fields, modelled fields, and MGEs) can be seen in Figure \ref{appendixfigure}. 

In general, the anisotropy and inclination of each galaxy were unconstrained in the \textsc{emcee} posterior distributions.  Both parameters were included to accurately reflect the uncertainty on the density slope and avoid driving the density slopes to solutions motivated by inaccurate anisotropy or inclination values. The priors were set such they span a realistic range of anisotropies as observed in local Universe studies \citep{cappellari_atlas3d_2013-1}. 

For some galaxies with a low number of spatial elements across the field (for example, galaxy 4439 in cluster A2744, in Figure \ref{appendixfigure}), the derived density slopes were well constrained in terms of uncertainties at the level of $ \sim15\% $.  This constraint is at face value surprising, as with few spatial elements, structure such as clear rotation or  comparatively high central dispersion is unobservable. However, as a global potential is set, only a small parameter space of break-radius densities and inner density slopes lead to an integrated mass that can replicate the observed kinematics. Even with few constraints, the dynamical mass found by JamPy is robust, as the mass directly impacts the observed kinematics, with no intermediate assumptions concerning the relative distributions of baryons and dark matter. As can be seen in Figure \ref{appendixfigure} columns 5 and 7, density slopes that are too steep or shallow lead to a mean $v_{\text{rms}}$ value that is too high or low, respectively.

\begin{figure}
  \begin{center}
      \includegraphics[width=\columnwidth]{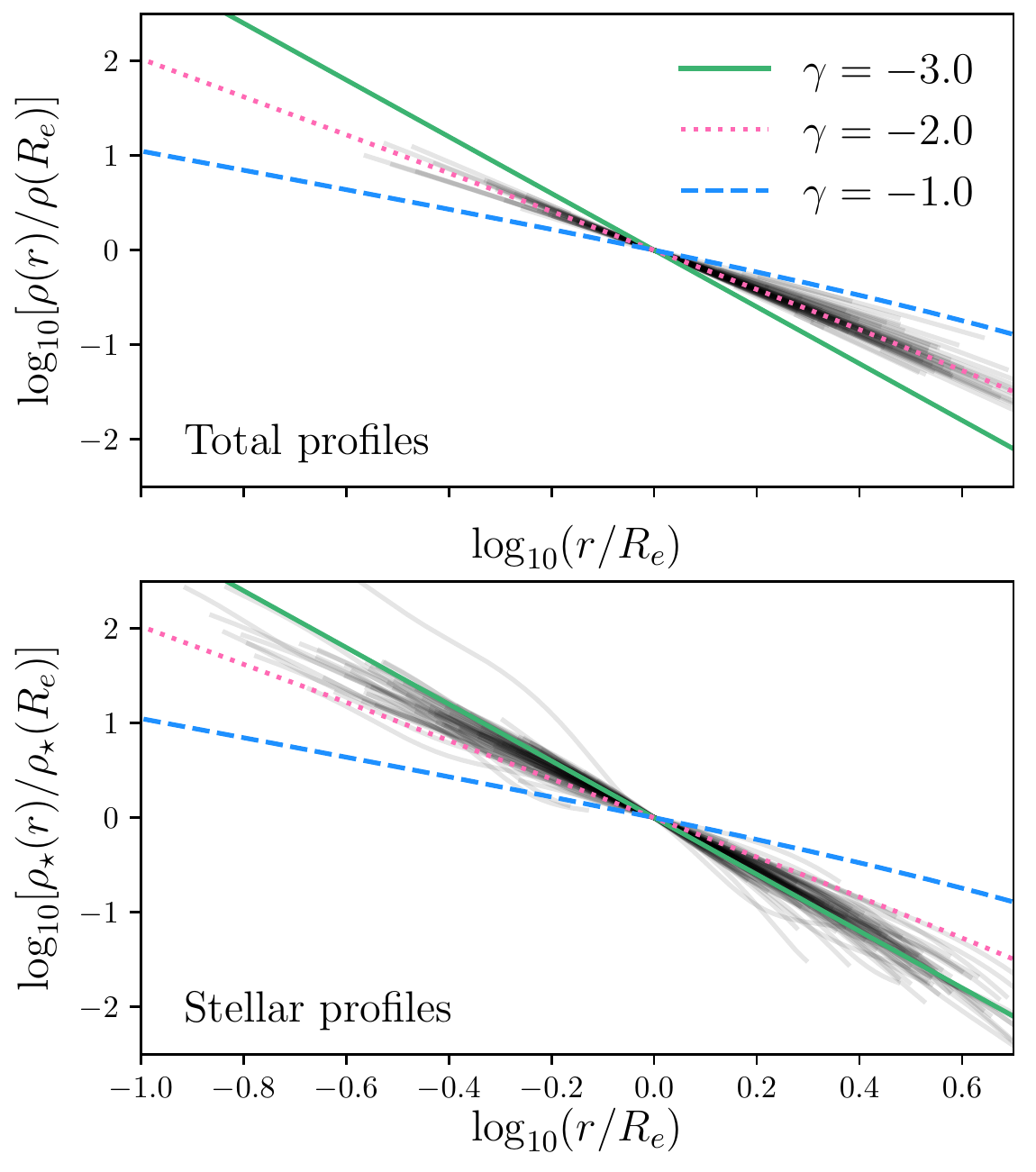}
      \caption{The Frontier Fields sample total density profiles are shown in the top panel, with isothermal and sub- and super-isothermal slopes shown for reference. The profiles are plotted between the kinematic MUSE psf radius for each cluster to the maximum kinematic data extent. These profiles are obtained from the models. The total profiles have density slopes around isothermal values. Stellar profiles measured directly from the stellar MGEs are shown on the bottom panel, between the radial range 0.16" to the maximum kinematics extent. The inner radius is chosen in both cases to avoid PSF effects of MUSE and the HST respectively. The stellar profiles are on average steeper than isothermal. }
      \label{allprofiles}
    \end{center}
\end{figure}

The derived total density profiles and the accompanying stellar density profiles are shown in Figure \ref{allprofiles}. The stellar profiles were computed directly from the stellar MGES fitted to the HST data, across the radial range of 0.16" to the maximum kinematic extent of the data. The inner radius is set so as to conservatively avoid the HST PSF. The total density slopes have a median of \g.  The distribution has a standard deviation of 0.21  and a median $1\sigma$ uncertainty on each slope of $0.11$. The stellar slopes are on average steeper than the total slopes, with a median of \s, and a standard deviation of 0.32. 

\subsection{Correlations with total mass density slope}
\label{corr}
\begin{figure}
  \begin{center}
    \includegraphics[width=\columnwidth]{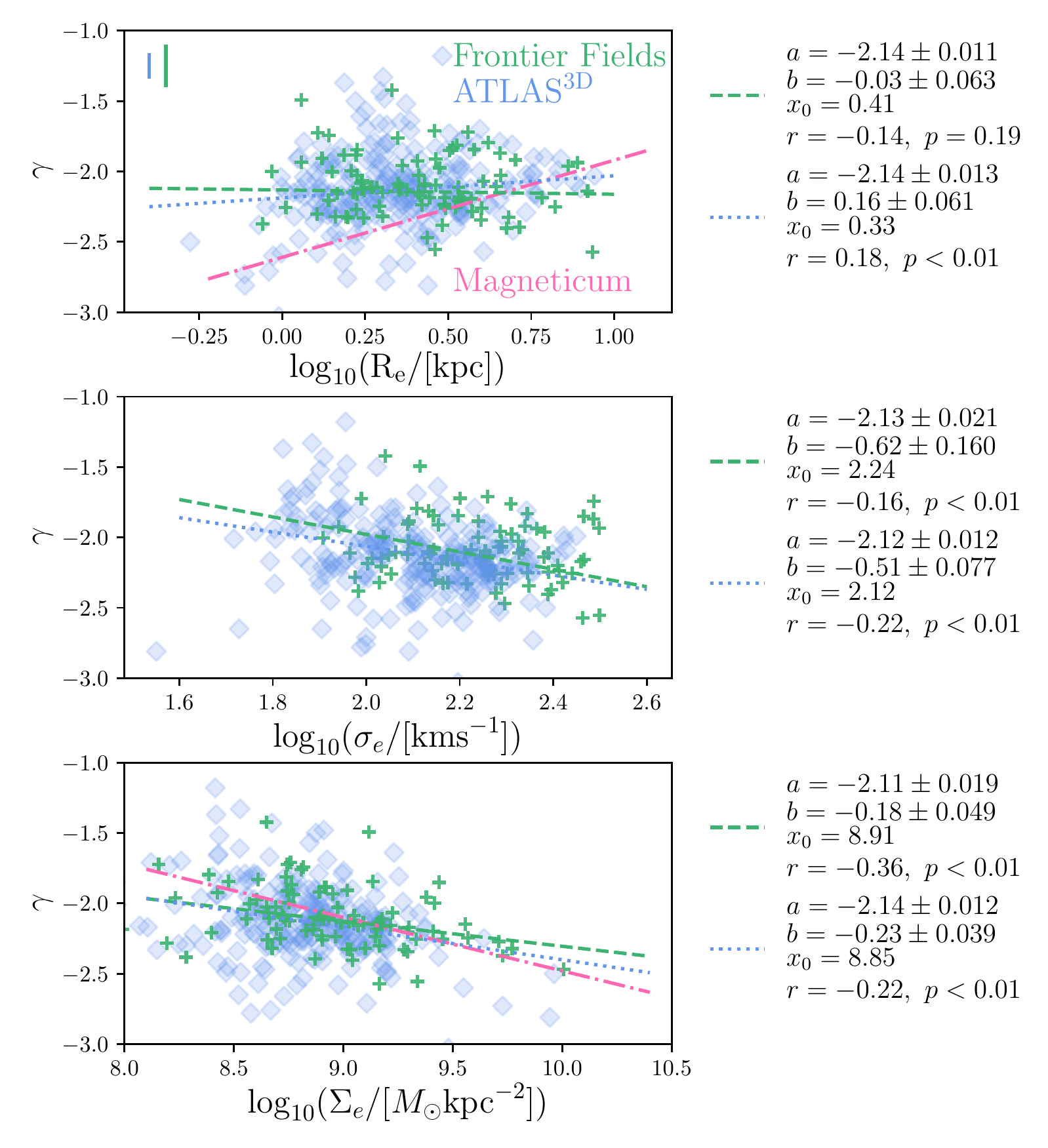}
      \caption{The observed relationships between density slope and effective radius (top panel), velocity dispersion (middle panel) and surface mass density (bottom panel) for the Frontier Fields sample (green) and $\text{ATLAS}^{\text{3D}}$ sample (blue). The green and blue dashed lines show the best fit obtained by the LtsFit procedure for the Frontier Fields and $\text{ATLAS}^{\text{3D}}$ samples, respectively, in the form $y = a + b (x - x_0)$. The best-fit parameters, corresponding errors, and Pearson coefficients with $p$-values are inset. A representative error on the density slope for each sample is shown in the top left corner of the top panel. The  $\text{ATLAS}^{\text{3D}}$ sample effective radii are from \citet{cappellari_atlas3d_2011}, with the velocity dispersions from \citet{cappellari_atlas3d_2013-1}, and slopes from \citet{poci_systematic_2017}. The LtsFit procedure was applied to the $\text{ATLAS}^{\text{3D}}$  sample in the same way as for the Frontier Fields galaxies. The Magneticum relation between density slope and effective radius is shown in the top panel, of form $\gamma = 0.69\log(\text{R}_{\text{e}} -2.61$). The Magneticum relation between density slope and surface mass density is shown on the bottom panel, of form $\gamma = -0.38\log(\Sigma_{\star}) +1.32$. The Magneticum fits are from \citet{remus_co-evolution_2017}.}
      \label{linear_fits}
    \end{center}
\end{figure}

The LtsFit\footnote{https://pypi.org/project/ltsfit/} procedure of \citet{cappellari_atlas3d_2013-1} was used to investigate correlations between the total density slope and structural properties in the Frontier Fields sample. All fits and underlying data can be seen in Figure \ref{linear_fits}.

For the Frontier Fields sample, no significant correlation with  (log) effective radius was found, consistent with the trend observed in the   $\text{ATLAS}^{\text{3D}}$ sample within 1$\sigma$. Comparing to trends from simulations, we see that the Magneticum simulation early-type galaxies show a significant correlation with effective radius, where galaxies with a smaller effective radius have on average steeper density slopes, with the linear fits remaining constant across different redshift ranges \citep{remus_co-evolution_2017}. The gradient of the Magneticum relation, ${d\gamma/d\text{R}_{\text{e}} = 0.69}$, is very similar to the redshift zero trend of the IllustrisTNG early-type galaxies of $d\gamma/d\text{R}_{\text{e}} = 0.64$. Using the SL2S, SLACS, and LSD samples, \citet{sonnenfeld_sl2s_2013} report a comparable trend of density slope with effective radius of ${d\gamma/d\text{R}_{\text{e}} = 0.76  \pm 0.15}$ for fixed redshift and mass, steeper than the SLACS only trend of ${d\gamma/d\text{R}_{\text{e}} = 0.41  \pm 0.12}$ found by \citet{auger_sloan_2010} for $ 0 < z < 0.36$. In summary, the observed trends between density slope and effective radius in the $\text{ATLAS}^{\text{3D}}$ and Frontier Fields samples are therefore not in agreement with the comparatively steep Magneticum, IllustrisTNG, and lensing relations. 

A trend between density slope and velocity dispersion is observed, seen in the middle panel of Figure \ref{linear_fits}. The Frontier Fields sample has a marginally steeper relation  with velocity dispersion ($d\gamma/d\sigma_\text{e} = -0.62$) than the $\text{ATLAS}^{\text{3D}}$ sample ($d\gamma/d\sigma_\text{e} = -0.51$), with high dispersion galaxies having steeper observed density slopes. However, given the scatter in the data, the recovered trends between samples are consistent within the measurement errors. No pronounced trend between central velocity dispersion and total density slope is found in the IllustrisTNG simulations for redshift zero \citep{wang_early-type_2020}, contrary to the relatively steep relation found here.

A near identical trend  between the Frontier Fields and  $\text{ATLAS}^{\text{3D}}$ samples was calculated for the relationship between total surface mass density and the total mass density slope, seen in the bottom panel of Figure \ref{linear_fits}. The Magneticum relation is also shown for the stellar mass surface density, which \citet{remus_co-evolution_2017} found to be redshift independent.  The surface density is defined as 
\begin{equation}
  \Sigma_{\text{e}} = \frac{0.5 \times M}{\pi R^2_{\text{e}}},
\end{equation}
\label{surfdeneq}
where the total mass, $\text{M}$, is twice the mass computed by integrating the best-fit total potential within a sphere of radius one R$_{\mathrm e}$, as in \citet{cappellari_atlas3d_2013-1}. The total masses used for this calculation are given in Table \ref{full_data}. Across both samples, objects with high surface mass densities have steeper slopes, indicating compact objects have correspondingly steep density profiles, as naively expected, with a gradient of $d\gamma/d\Sigma_{\text{e}} = -0.18$ for the Frontier Fields sample, and a gradient of $d\gamma/d\Sigma_{\text{e}} = -0.23$ for the $\text{ATLAS}^{\text{3D}}$ sample.  \citet{sonnenfeld_sl2s_2013} report a trend of $\delta \gamma/\delta \log \Sigma_{\star} = - 0.38 \pm 0.07$ for gravitational lensing results, identical to the Magneticum relation for early-type galaxies \citep{remus_co-evolution_2017}, both using a stellar surface mass density instead of a total surface mass density. The steepest relation is given for the IllustrisTNG early-type galaxies, of $d\gamma/d\Sigma_{\star}= -0.45 \pm 0.02$ at $z = 0$ \citep{wang_early-type_2020}. 

The agreement of the above trends in the $\text{ATLAS}^{\text{3D}}$ and Frontier Fields samples,  despite the different redshifts, indicates the dependence of the density slope on these structural parameters is independent of redshift. This is in agreement with the results from \citet{remus_co-evolution_2017}, who also found the correlations with stellar surface mass density and effective radius to be independent of redshift for the Magneticum early-type galaxy sample.

\begin{figure*}
  \begin{center}
      \includegraphics[width=2\columnwidth]{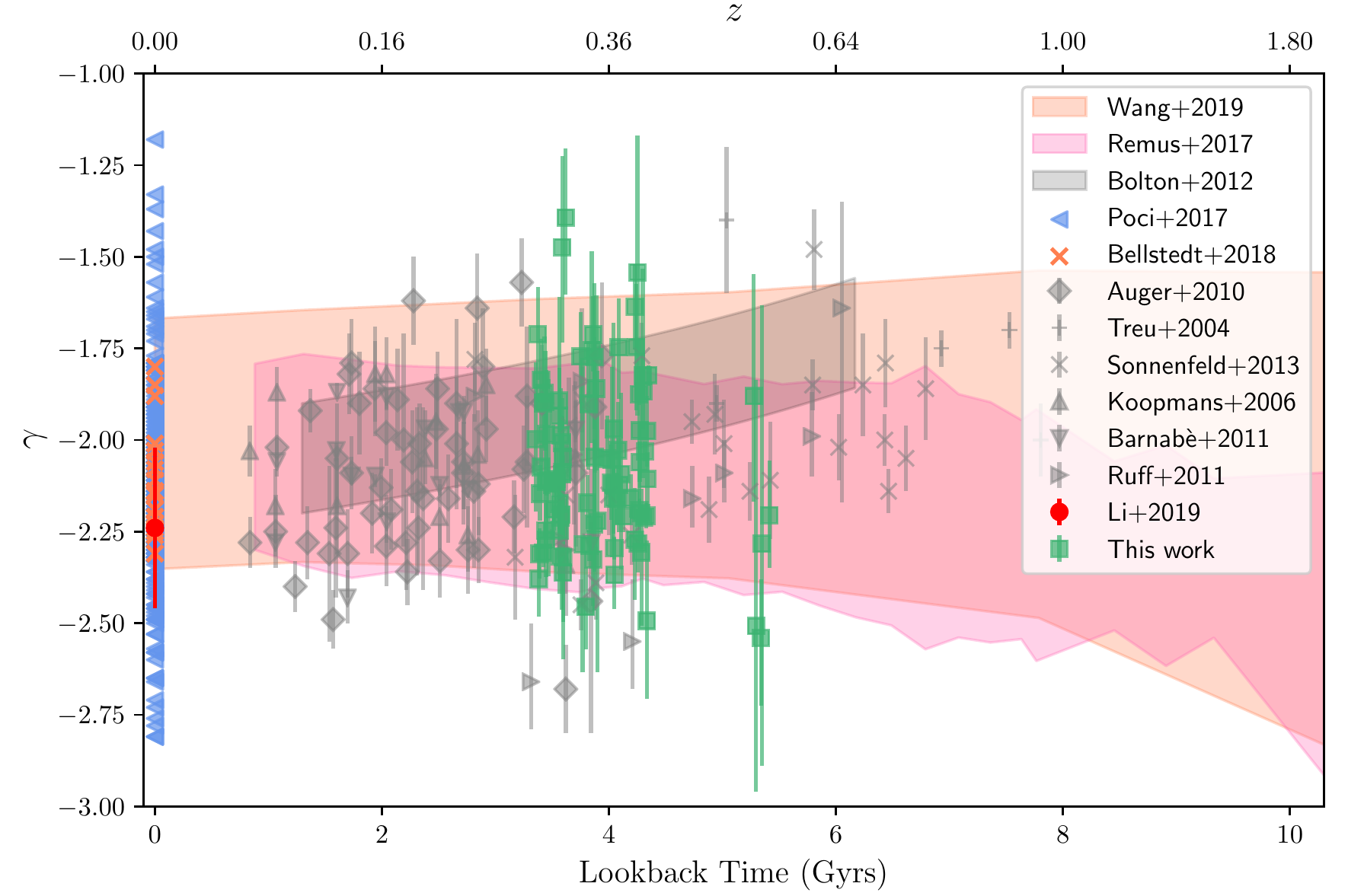}
      \caption{Measured density slopes using different methods are shown. At redshift zero, for context, are the results of \citet{poci_systematic_2017} Model I, which is an identical model construction to the generalised potential used in this work. The IllustrisTNG results, from \citet{wang_early-type_2019-1} are across seven redshifts bins (0,0.1,0.2,0.3,0.5,1,2), and the results are shown as an orange region, of width equal to the twice standard deviation at each redshift. Magneticum simulation results are shown as a pink shaded region, defined here by a parabolic fit to the model predictions, with width equal to twice the standard deviation of the density slopes at each redshift. All gravitational lensing results are shown in grey: see \citet{treu_massive_2004,koopmans_sloan_2006,auger_sloan_2010,ruff_sl2s_2011,barnabe_two-dimensional_2011,sonnenfeld_sl2s_2013}. The density slope evolution with redshift measured by \citep{bolton_boss_2012} is shown as a black shaded band, with width corresponding to the quoted uncertainty. The derived density slopes for this work are shown in green with their $1\sigma$ uncertainty, corresponding to the 16th and 84th percentiles of the \textsc{emcee} distribution.}
      \label{evo}
  \end{center}
\end{figure*}

 All density slope values for the Frontier Fields sample are shown against redshift in Figure \ref{evo}. No significant correlation is observed between density slopes in the Frontier Fields sample and redshift in the range $0.29 < z < 0.55$. However, strong evidence for evolution in the small redshift range of the Frontier Fields sample is not expected, considering also the intrinsic scatter of the sample and associated density slope uncertainties. The IllustrisTNG simulations indicate little evolution in density slope below $z \sim 1$, due to evolution via gas poor mergers \citep{wang_early-type_2019-1}. The Magneticum simulations predict a mild trend with redshift given by the relation $\langle\gamma\rangle = -0.21z - 2.03$ \citep{remus_co-evolution_2017}. Gravitational lensing indicates density slopes were shallow at high redshifts, and have progressively become steeper as redshift approaches zero, with the most pronounced relation being  $d\langle\gamma\rangle/dz = 0.60 \pm 0.15$ \citep{bolton_boss_2012} for the redshift range $ 0.1 < z < 0.7$. Using the SL2S, SLACS, and LSD samples, \citet{ruff_sl2s_2011} finds a milder redshift trend of $d\langle\gamma\rangle/dz = 0.25 ^{+0.10}_{-0.12}$, up to redshifts of $z \sim 1$. A similar analysis presented by \citet{sonnenfeld_sl2s_2013} with 25 lensing systems from SL2S found a redshift trend $d\langle\gamma\rangle/dz = 0.31 \pm 0.10$, for $ 0.2 < z < 0.8$.  There is no evidence of shallower slopes at greater lookback times in the Frontier Fields sample, contrary to the results of gravitational lensing. 

\subsection{Comparison to simulations at $0.29 < z < 0.55$}

\begin{figure}
  \begin{center}
      \includegraphics[width=\columnwidth]{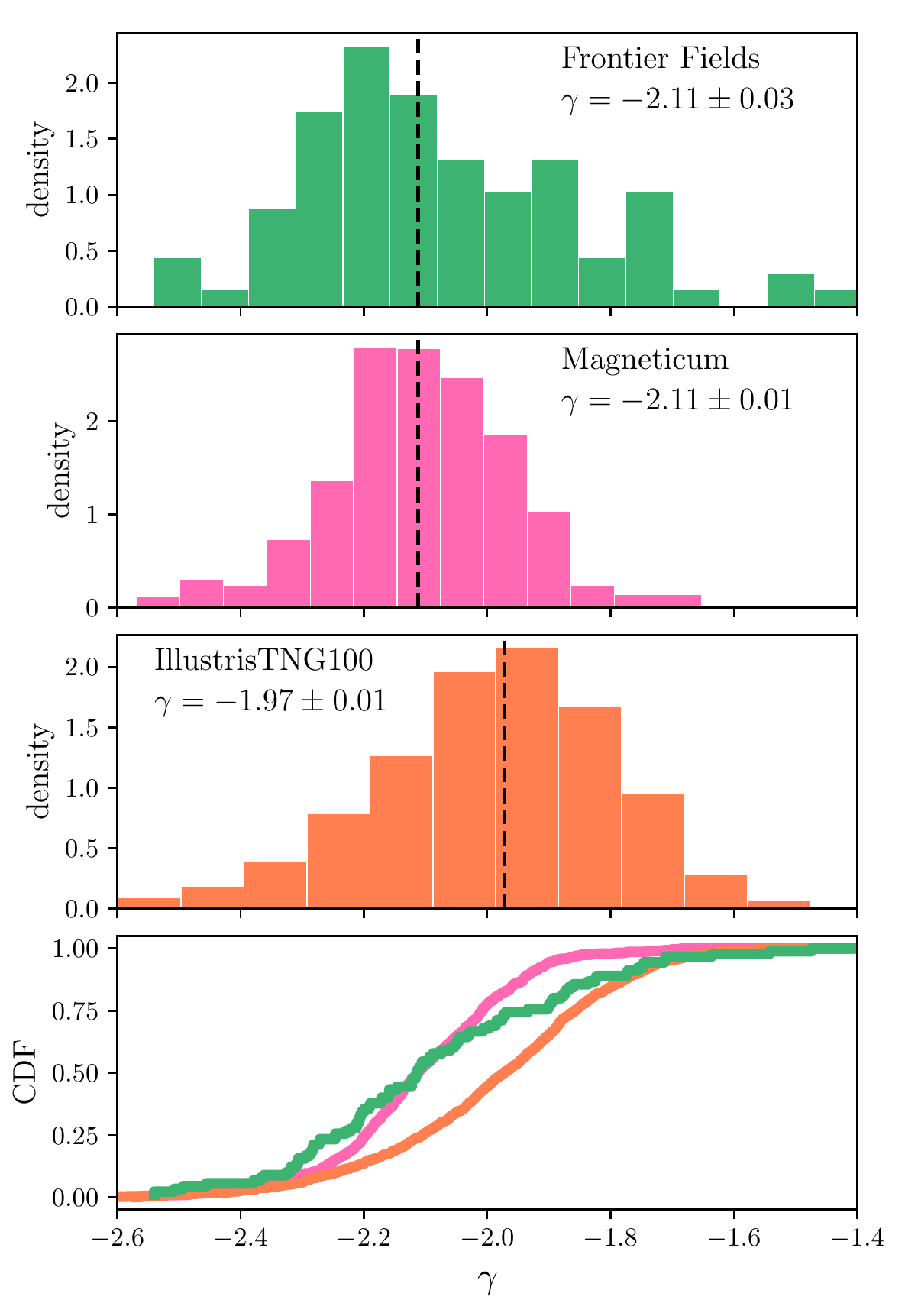}
      \caption{The Frontier Fields sample density slopes (green) compared to the Magneticum simulation galaxy density slopes (pink) and IllustrisTNG density slopes (orange). The histograms are plotted as density distributions instead of frequency. Inset are the median density slope and associated standard error. The bottom panel shows the cumulative distribution function (CDF) of each sample, which forms the basis of the KS-tests.}
      \label{simhist}
    \end{center}
\end{figure}

The distribution of total mass density slopes of the Frontier Fields sample was compared to the distribution of total mass density slopes calculated from early-type galaxies in the Magneticum simulations \citep{teklu_connecting_2015,remus_co-evolution_2017} and early type galaxies in the  IllustrisTNG100 simulations \citep{pillepich_simulating_2018,springel_first_2018}, shown in Figure \ref{simhist}. The cumulative distribution of the samples is shown in the bottom panel.  To make this comparison, only Magneticum early-type galaxies (724 galaxies) in the redshift range $0.29 < z <  0.55$ were used, and early-type IllustrisTNG  galaxies in two redshift bins ($z= 0.3$ and $z = 0.5$,  1432 galaxies).

The total mass density slopes of the Frontier Fields sample at this redshift agree well with the predictions of the Magneticum simulations in terms of their medians. A Kolmogorov-Smirnov (KS) two-sample test performed on the Magneticum and Frontier Fields density slopes yields a KS test statistic of 0.13, and a critical value of 0.18 for $\alpha = 0.01$. The obtained p-value is 0.10, failing to reject the null hypothesis the samples share a common distribution. The Magneticum simulation prediction of density slopes for early-type galaxies are therefore consistent with the observed Frontier Fields galaxy slopes in the same redshift range. 

The IllustrisTNG sample has on average shallower slopes for the same redshift, with a skew distribution towards steeper slopes, indicating a predominance of more compact objects in that sample. The IllustrisTNG early-type galaxy density slopes are not consistent with the Frontier Fields sample slopes, with the medians differing by more than $3\sigma$. A KS test for the two samples gives a test statistic of 0.35, a critical value of 0.1769, and a p-value of $7 \times 10^{-10}$, indicating the IllustrisTNG and Frontier Fields sample do not share a common distribution.  The shallower IllustrisTNG slopes may arise due to the level of included active-galactic nuclei (AGN) feedback in the models, which quenches in-situ star formation, removes baryonic matter from central regions, and causes density slopes to approach isothermal values \citep{wang_early-type_2019-1,wang_early-type_2020}. In particular, \citet{wang_early-type_2019-1} find AGN feedback energy through the kinetic mode (the ejection of kinetic energy and momentum into surrounding gas cells without thermal energy) as implemented in the IllustrisTNG simulations dominates the change of density slopes towards shallower values in the redshift range $ 1 < z< 2$, after which the density slope is near invariant. 

\subsection{Comparison to local Universe density slopes}

The derived total mass density slopes in comparison to other studies, as a function of redshift, are shown in Figure \ref{evo}. A unique aspect of this work is that we apply exactly the exact same methodology to the Frontier Fields sample as applied to the local $\text{ATLAS}^{\text{3D}}$ sample, allowing the two sets of density slope values to be directly compared. In particular, the comparison was made using the `Model I' results of \citet{poci_systematic_2017}, where the same generalised potential is used, in the form of a spherical double power law. A point of difference between the samples is the radial range, which extends to 3\re\, on average for the Frontier Fields galaxies, and 1\re\, for the $\text{ATLAS}^{\text{3D}}$ sample. However, \citet{bellstedt_sluggs_2018} find a comparable mean density slope using a subset of the $\text{ATLAS}^{\text{3D}}$ sample with the dynamical models constrained up to 4\re, signifying the differences in radial range between the $\text{ATLAS}^{\text{3D}}$ sample and Frontier Fields sample is not problematic. Furthermore, \citet{cappellari_small_2015} find a power law with constant logarithmic slope $\gamma$ is an accurate description of the total mass density slope across the radial range 0.1 - 4 \re. 

\begin{figure}
  \begin{center}
      \includegraphics[width=\columnwidth]{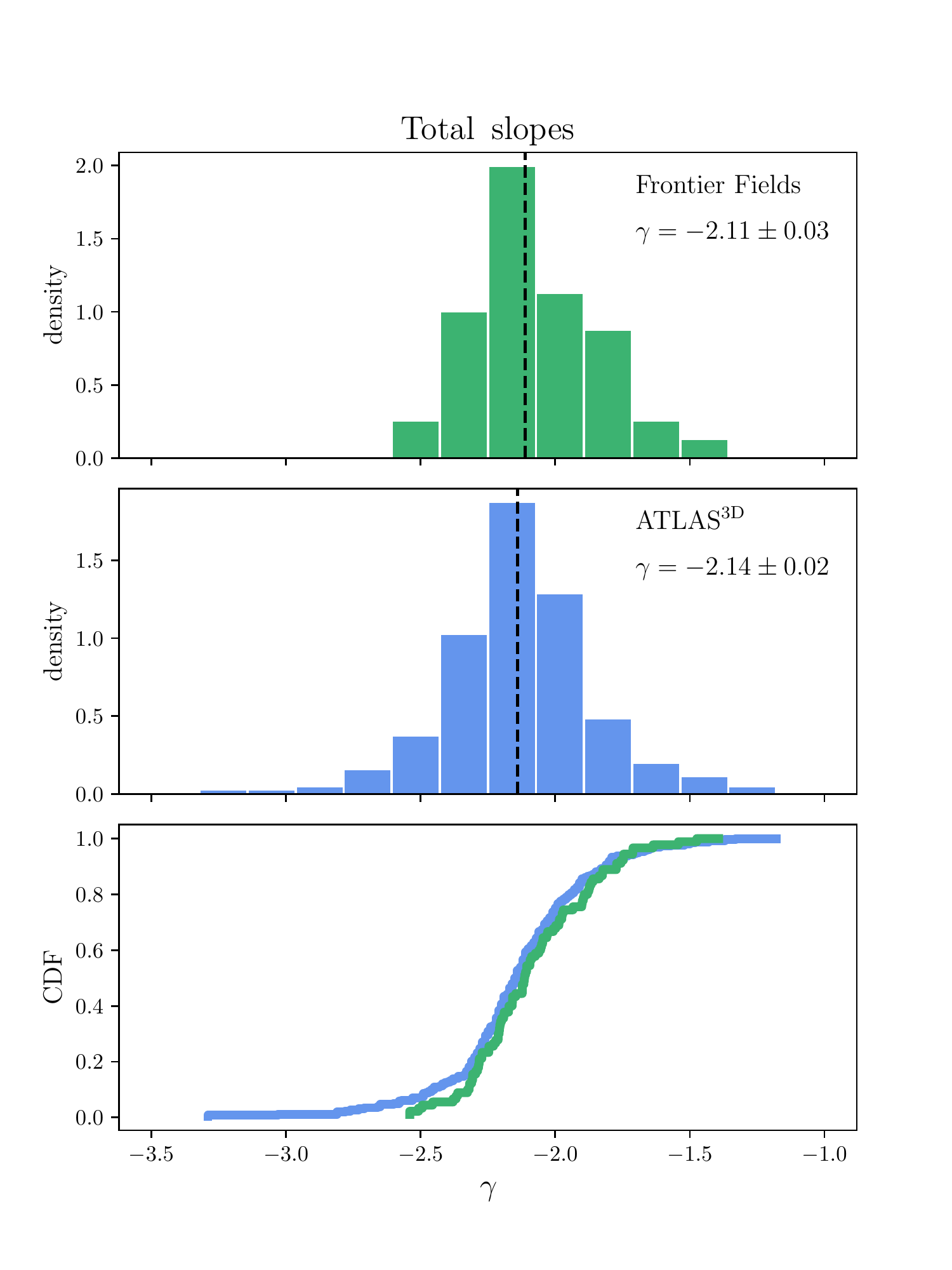}
      \caption{The top panel (green) shows the density histogram of the Frontier Fields sample with the median density slope and standard error inset. The lower panel (blue) shows the density histogram of the $\text{ATLAS}^{\text{3D}}$ sample of \citet{poci_systematic_2017} at redshift zero, with the median slope and standard error inset. The cumulative distribution of the two samples is shown on the bottom panel.}
      \label{obhist}
    \end{center}
\end{figure}

The distribution of density slopes for the 258 galaxies of the $\text{ATLAS}^{\text{3D}}$ sample is compared to the Frontier Fields sample in Figure \ref{obhist}. The $\text{ATLAS}^{\text{3D}}$ sample has a median value and standard error of $\gamma = -2.14 \pm 0.02$, and the Frontier Fields sample of \g. This result is in good agreement with the results of \citet{bellstedt_sluggs_2018}, who found $\langle\gamma\rangle = -2.12 \pm 0.05$ for 21 early-type galaxies, but is shallower than the ManGA mean density slope of $\langle\gamma\rangle = -2.24$ \citep{li_sdss-iv_2019}. However, the ManGA sample is not restricted to early-type galaxies. 

The $\text{ATLAS}^{\text{3D}}$ sample has a tail of steeper density slopes that are not observed in the Frontier Fields sample, seen in Figure \ref{obhist}. Given the correlations discussed in Section \ref{corr}, this tail represents compact galaxies with high surface mass densities. A tail of shallow density slopes is also missing from the Frontier Fields sample, which could be an effect of extended, low surface brightness galaxies being difficult to observe at large distances. It is possible the dense cluster environment of the Frontier Fields sample also had an effect. However, no statistical difference was found between the subsample of  $\text{ATLAS}^{\text{3D}}$ drawn from the Virgo cluster and the $\text{ATLAS}^{\text{3D}}$ sample as a whole, suggesting environmental differences do not significantly impact density slope distributions. Note the comparison of total mass density slopes is drawn against the full $\text{ATLAS}^{\text{3D}}$ sample of galaxies. A break has been reported in the $\gamma-\sigma_{\text{e}}$ relation, where the density slope depends on the velocity dispersion of galaxies with a central dispersion below $\sim 100 \text{ kms}^{-1}$. Density  slopes are found to be independent of velocity dispersion for values above $\sim 100 \text{kms}^{-1}$ \citep{poci_systematic_2017,li_sdss-iv_2019}. The agreement between the Frontier Fields sample and the $\text{ATLAS}^{\text{3D}}$ sample is not impacted by only considering galaxies for which $\sigma_{\text{e}} > 100 \text{ kms}^{-1}$.

A KS test was performed between the $\text{ATLAS}^{\text{3D}}$ and Frontier Fields samples, with the cumulative distribution function shown in the bottom panel of Figure \ref{obhist}. A KS test statistic of 0.10 was obtained for a critical value of 0.19, with $\alpha = 0.01$, and a p-value of 0.47. The null hypothesis that the samples share a common distribution is not rejected. As the same methods were used to derive the density slopes for the $\text{ATLAS}^{\text{3D}}$ and Frontier Fields sample, the indication of the KS statistic that the samples were drawn from comparable distributions reveals no evolution of the total density slope in the last $\sim 6$ Gyrs of cosmic time.

From the MGE fits to the HST photometry, stellar density slopes are also derived for the Frontier Fields, defined here as 
\begin{equation}
  \label{gammstaravg}
  \gamma_{\star} = \frac{\log_{10}\left(\rho_{\star}(R_{\text{o}})/\rho_{\star}(R_{\text{i}})\right)}{\log_{10}\left(R_{\text{o}}/R_{\text{i}}\right)},
\end{equation}
with $\rho_{\star}$ the stellar density inferred from the MGEs, and $R_{\text{i}}$ and $R_{\text{o}}$ the inner and outer radius, respectively. 
The same method is applied to the $\text{ATLAS}^{\text{3D}}$ galaxies using the MGEs of \citet{scott_atlas3d_2013-1}. However, due to the differences in PSF sizes and the spatial coverage of the MGEs, the Frontier Fields stellar slopes are computed across the radial range 0.16" to the maximum kinematic extent of the data (on average 3\re), while the $\text{ATLAS}^{\text{3D}}$ stellar slopes are computed across the radial range 2" to the maximum kinematic extent (on average 1\re).

\begin{figure}
  \begin{center}
      \includegraphics[width=\columnwidth]{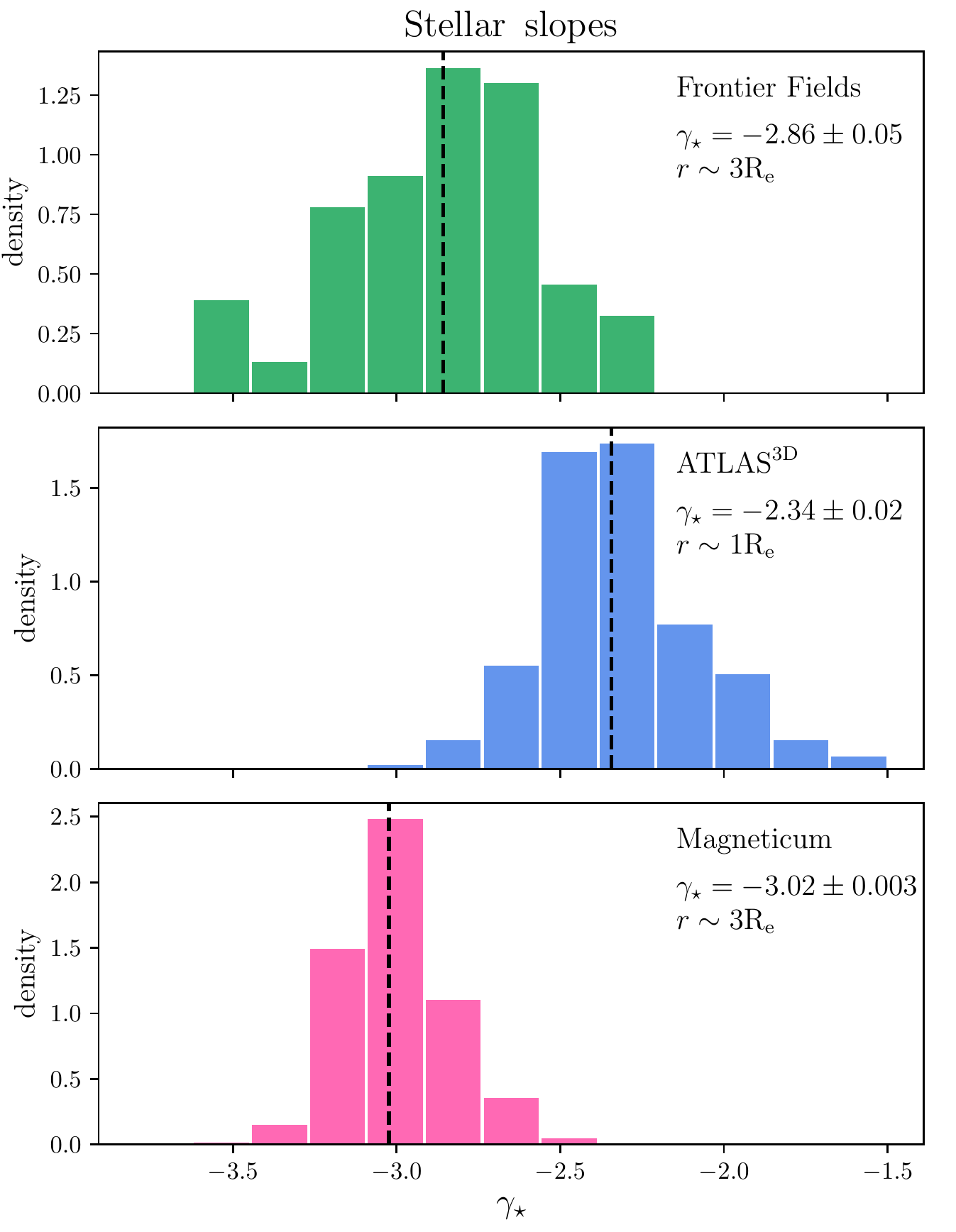}
      \caption{The Frontier Fields sample stellar density slopes (green) as measured from the stellar MGEs, compared to the  stellar density slopes of the $\mathrm{ATLAS}^{3\text{D}}$ sample (blue), measured from the MGEs of \citet{scott_atlas3d_2013-1}. The median value and its associated standard error is shown for each sample. The $\mathrm{ATLAS}^{3\text{D}}$ stellar slopes are calculated up to 1\re, while the Frontier Fields sample is calculated to 3\re\,on average. The Magneticum slopes shown on the bottom panel (pink) are calculated up to 3\re, and are shown for the redshift range of the Frontier Fields sample. The difference in radial ranges prevents a robust comparison between the distributions.}
      \label{hist_stars}
    \end{center}
\end{figure}

The stellar slopes found for the Frontier Fields sample, with median \s, are steeper than the $\text{ATLAS}^{\text{3D}}$ stellar slopes of median $\gamma_{\star} = - 2.34 \pm 0.02$. The distribution are shown in Figure \ref{hist_stars}. However, as noted in \citet{poci_systematic_2017}, the stellar profiles become steeper with increasing radius, and so the difference in radial ranges between the two samples becomes critical here. That the stellar slopes are steeper than the total density slopes for both samples is in agreement with simulations, which shows that the total slope is dominated by the stellar profile at small radii and the dark matter profile at large radii \citep{remus_dark_2013}. In addition, the stellar profiles are not well described by a simple power law, but are more curved and thus the power law slopes become increasingly steep for larger fitted radii. In particular, the Magneticum slopes and Frontier Fields stellar slopes extend to larger radii than the $\text{ATLAS}^{\text{3D}}$ slopes, and have correspondingly steeper values. The Magneticum slopes have a median of ${\gamma_{\star} = -3.06 \pm 0.003}$ for the redshift range of the Frontier Fields sample. The profiles cannot be computed across a common radial range due the PSF effects in the inner regions for the Frontier Fields sample,  the softening length in the Magneticum simulations which prevents the stellar slopes from being inferred at small radii, and the limited radial extent of the $\text{ATLAS}^{\text{3D}}$ profiles.

The relative spread of the distributions between stellar slopes and total mass density slopes can be used to investigate the so called  `bulge-halo conspiracy', where steep light profiles combine with dark matter profiles to form an isothermal total profile \citep{dutton_bulgehalo_2014}. \citet{poci_systematic_2017} found that the scatter in the total profiles was only marginally smaller than the stellar profiles for the $\text{ATLAS}^{\text{3D}}$ sample, and as such provided little evidence for a `conspiracy'. Here, we note that the sensitive dependence of the stellar slopes on the precise radial range used and the limitations of assuming a power law naturally cause the stellar slope values to have larger scatter than the total slope values, which should not be read as evidence towards a bulge-halo conspiracy. 

\section{Discussion}
\label{discussion}
Assuming that the selected sample of Frontier Fields galaxies is representative of the local $\text{ATLAS}^{\text{3D}}$ sample, the similarity of the density slope distributions indicates there has been no change in the average density slope in the past $\sim 6$ Gyrs of cosmic time. The consistency of the density slope across this span of time points to an evolutionary process for early-type galaxies which does not significantly perturb the total density slope, such as the dry merging phase described in the two-phase model. 

In the two-phase model, galaxies grow in mass through gas-rich accretion events at high redshifts, causing dense central regions and significant rates of star formation. At lower redshifts, relatively frequent gas-poor mergers increase the size of the host galaxy without substantially increasing the mass \citep{naab_minor_2009,nipoti_dry_2009-1,hilz_relaxation_2012}. The accretion events at large radii also leave the central regions of the host galaxy unperturbed \citep{karademir_outer_2019}. As a consequence, the galaxy `puffs-up', resulting in a shallower total mass density slope. \citet{remus_dark_2013} note that the gas-poor nature of the merger is necessary to drive total mass density slopes towards isothermal values of $\gamma \sim -2$, and that major mergers are also capable of causing this evolution. However, \citet{mclure_sizes_2013} note that unfeasible rates of major mergers would be required to produce local passive galaxy characteristics from their higher redshift counterparts, and would result in an over-population of high mass stars in the local stellar mass function. \citet{mclure_sizes_2013} therefore propose a minor-merger evolutionary pathway for passive galaxies, where passive indicates the absence of in-situ star formation. 

It is interesting that dry merger events cause density slopes to approach isothermal values. \citet{remus_dark_2013} argue that approximately isothermal solutions for the total density profile are an `attractor', as once a galaxy has reached an isothermal density slope, subsequent dry merger events will not change it. Total density slopes might tend towards isothermal solutions due to an incomplete violent relaxation process, such as described by \citet{lynden-bell_statistical_1967}. Only merger events with high gas fractions are capable of perturbing isothermal density profiles to steeper values again \citep{remus_dark_2013}. A similar behaviour is seen in the simulations of \citet{nipoti_dry_2009-1}, where dry mergers preserve the isothermal structure of galaxies. The lack of observed density slope evolution between the Frontier Fields sample and $\mathrm{ATLAS}^{3\text{D}}$ sample provides evidence for approximately isothermal density slopes acting as an `attractor'. This is consistent with the Magneticum and IllustrisTNG simulations, which show only mild evolution in the total density slope in the past 6 Gyrs of cosmic time \citep{remus_co-evolution_2017,wang_early-type_2019-1}. 

The Magneticum simulations quantify the mild evolution in the total density slope with redshift, of the form $d\gamma/dz = -0.21z - 2.03$, so that early-type galaxies at higher redshifts have on average steeper slopes \citep{remus_co-evolution_2017}. The IllustrisTNG simulations, with shallower slopes on average, indicate almost no evolution in the density slope below $z \sim 1$ \citep{wang_early-type_2019-1}. In this regard, the fact the $\mathrm{ATLAS}^{3\text{D}}$ and Frontier Fields density slopes have comparable distributions is more in line with the IllustrisTNG predictions. To further test the correlation between total density slope and redshift as predicted by different simulations, it is necessary to extend the redshift baseline. 

A direct, statistical comparison is not made between the Frontier Fields slopes and gravitational lensing slopes, as there are not enough data points within the same redshift range to make a meaningful comparison. However, the lack of evolution with redshift between the Frontier Fields and $\mathrm{ATLAS}^{3\text{D}}$ samples is in tension with the correlation with redshift of the lensing samples, found by \citet{ruff_sl2s_2011}, \citet{bolton_boss_2012}, and \citet{sonnenfeld_sl2s_2013}. Given the lensing trend of shallower density slopes at greater redshifts, \citet{sonnenfeld_purely_2014} argue there must be continued accretion of modest but non negligible amounts of gas during dry merger events. The steeper density slopes at low redshift result from the gas condensing in the centre of the galactic potential, leading to small amounts of star formation and denser central regions, with little change to the galactic radius. However, too much cold gas accretion leads to adiabatic contraction, and a reduction in effective radius.  A contracting radius with redshift is at odds with the expected redshift-size growth of early-type galaxies, which have been shown observationally to increase in agreement with cosmological models \citep{khochfar_simple_2006,trujillo_strong_2007,huertas-company_evolution_2013,van_der_wel_3d-hstcandels_2014}. In addition, the ongoing star formation  contradicts studies showing early-type galaxies have predominately old and uniform stellar populations \citep{thomas_epochs_2005}.

The two-phase model is however consistent with the quiescent stellar populations that characterise early-type galaxies. Stellar populations form in place at early times, with a so-called `frosting' of younger stars added to the underlying older populations at later times \citep{trager_stellar_2000,diaz-garcia_stellar_2019}. \citet{mcdermid_atlas3d_2015} found star formation ceased earlier in the Virgo cluster environment compared to the field sample from a star formation history analysis, providing further evidence against ongoing gas accretion at late times in dense environments like that of the Frontier Fields galaxies. Recent studies of galaxy kinematics at high redshifts ($0.6 < z< 1 $)  also support a dry merger mass build-up scenario for early-type galaxies, restricted to a stable scaling relation between the half-light radius, velocity dispersion, and surface brightness, known as the mass fundamental plane \citep{graaff_tightly_2020}.

 \citet{bolton_boss_2012} instead suggest  the importance of off-axis major mergers over continued gas accretion in order to produce the lensing observations of  total density slopes that steepen as redshift approaches zero. However, \citet{newman_can_2012} report the size-growth of quiescent galaxies, as measured from observations, is a factor of $\sim 3.5$ in the redshift range $0.4 < z < 2.5$, which cannot be explained by the comparatively rare occurrence of major mergers as found by \citet{bundy_greater_2009}. Furthermore, \citet{newman_can_2012} find that dry mergers, with their high growth efficiency, can drive most or all of the  observed size-growth of quiescent galaxies for $ z < 1 $. 

Gravitational lensing studies at lower redshifts, such as those of \citet{auger_sloan_2010}, \citet{barnabe_two-dimensional_2011}, and \citet{shajib_dark_2021}, with samples at $z \sim 0.2$,  are consistent with the mean density slopes of this work, with mean density slope measurements of $\langle\gamma\rangle \sim -2.1$. \citet{shajib_dark_2021} suggest the measured slopes from lensing at $z \sim 0.2$, combined with correlations between effective radius and central dark matter fractions, are indicative of dry  merging driving the growth of early-type galaxies at low redshifts.  It  is the lensing measurements of density slopes at higher redshifts ($z \sim 0.6  - 1$) which indicate shallower mean slopes at greater lookback times.

When comparing the density slope distributions, it is also important to note the differing environments of the Frontier Fields and $\text{ATLAS}^{\text{3D}}$ galaxies.  Some studies indicate that cluster environments accelerate the galaxy evolution process, leading to compact, passive galaxies \citep{poggianti_evolution_2013-1}, although this is debated \citep{huertas-company_evolution_2013,shankar_size_2013}. \citet{gallazzi_galaxy_2021} find massive, passive satellite galaxies in haloes have a systematic excess in their light-weighted stellar ages and metallicities compared to central galaxies, indicating galaxies with older stellar populations prefer high density regions.  At present, there is no clear evidence environment significantly impacts the density slope measurement. 

It is worth noting that the density slopes as predicted by cosmological simulations are derived via a different method to this work.  While the Frontier Fields total density slopes are found through dynamical modelling, the simulation slopes are found by co-adding simulation particles within concentric, spherical shells, and fitting with a power law. The impact of the different methods is not yet clear, although \citet{remus_co-evolution_2017} note the discrepancy between gravitational lensing density slopes and those from the Magneticum simulations disappear if lensing techniques are applied to the simulation data. It has however been shown that the JAM method is robust for calculating dynamical masses of galaxies, by comparing JAM derived masses and known masses from simulated galaxies in the Illustris project \citep{li_assessing_2016}. 

The agreement between the Frontier Fields and $\text{ATLAS}^{\text{3D}}$ density slope distributions is interpreted as support for the two-phase model of early-type galaxy evolution, with no observed trend of shallower slopes at greater lookback times as found by some lensing works. Extending the study of spatially-resolved stellar kinematics to representative samples of higher-redshift galaxies, using consistent and homogenous modelling techniques such as that presented here, is required to resolve the apparent tension in galaxy mass density profiles coming from lensing, dynamics, and simulation predictions.

\section{Conclusions}
\label{concs}
We present the largest systematic study of two-dimensional stellar kinematic of galaxies at $0.29 < z < 0.55$ to date, using identical general dynamical modelling techniques as employed in large local studies. A sample of 90 galaxies from the Frontier Fields project was analysed using two-dimensional kinematics fields, HST photometry, and Jeans dynamical modelling to obtain total mass density slopes in the redshift range $0.29 < z < 0.55$. The main results are as follows: 

\begin{enumerate}
  \item The median total mass density slope obtained is \g\, (standard error). The distribution has a standard deviation of 0.23 and a median $1\sigma$ uncertainty on each slope of $0.11$.
  \item The obtained density slopes in the Frontier Fields sample agree well with the simulated Magneticum early-type galaxies in the same redshift range. The Frontier Fields density slopes are on average steeper than those of the IllustrisTNG early-type galaxies.
  \item The distribution of slopes obtained in this work is in agreement with slope distribution in the redshift zero study by \citet{poci_systematic_2017} using $\mathrm{ATLAS}^{3\text{D}}$ galaxies. The agreement indicates there has been no evolution of the total mass density slope in the last 4-6 Gyrs of cosmic time. 
  \item The Magneticum simulations predict a mild evolution in the total density slope in the past 6 Gyrs of cosmic time towards shallower slopes. This evolution is not observed when comparing the $\mathrm{ATLAS}^{3\text{D}}$ and Frontier Fields slope distributions, although pushing to higher redshift will be important to further test this. The lack of evolution in slope is more in agreement with the IllustrisTNG predictions, where the density slope is approximately constant for $z < 1$. 
  \item Trends between the total mass density slope and effective radius, velocity dispersion, and surface mass density were investigated. No significant correlation was found with effective radius. However, as for \citet{poci_systematic_2017}, it was found that galaxies with higher velocity dispersions, and those with higher surface mass densities,  have correspondingly steeper slopes. That the found relations are comparable to those found in \citet{poci_systematic_2017} indicate similar sample properties and no evolution in the scaling relations. 
  \item The lack of evolution in the density slope is not consistent with observations of density slopes using gravitational lensing methods, which indicate density slopes were shallow at greater lookback times and have become progressively steeper.
\end{enumerate}

The above results place some of the first constraints on the evolution of galaxy density slopes from a consistent application of stellar dynamics to two-dimensional kinematic data, complementing the findings so far only obtainable from gravitational lensing for significant samples. While this reveals elements of agreement and tension with both existing observations and simulations, upcoming developments will help improve the precision of these findings, in particular around the influence of environment, and extending the redshift baseline within which spatially-resolved kinematics are available.

New observations  of early-type galaxies across a span of redshifts  will be aided  by the now operational  ground-layer adaptive optics system of MUSE, which will alleviate the issue of relatively poor kinematic data quality for intermediate redshifts ($0.2 < z< 0.5$); such an application is the new MAGPI\footnote{https://magpisurvey.org} survey \citep{foster_2021_magpi}. The MAGPI survey has secured 340 hours of MUSE time, to obtain deep, adaptive-optics assisted observations of 160 relatively isolated galaxies at redshifts of $0.25 <z < 0.35$, avoiding the issues of crowding, limited resolution and low SNR encountered here. The MAGPI survey will help constrain the picture of galaxy evolution in the middles ages of the Universe, and will also help answer questions of the impact of environment on the density slope. 

Even further ahead, the James Webb Space Telescope will be able to deliver kpc-scale stellar kinematics at $z=1$ and higher. Planned instruments like MAVIS on the VLT \citep[][Rigaut et al. 2021, in prep]{mcdermid_phase_2020} will be able to provide sub-kpc resolution integral field unit stellar kinematics out to $z=1$, with HARMONI on E-ELT giving comparable resolution at redshifts beyond 1. Therefore, there will exist a slew of instruments which will deliver high quality kinematic data, allowing dynamical modelling to probe even higher redshifts.

\section*{Acknowledgements}
We thank the referee Oliver Czoske for the many helpful suggestions that improved this article.
This work is based on observations made with the NASA/ESA Hubble Space Telescope, obtained from the data archive at the Space Telescope Science Institute. STScI is operated by the Association of Universities for Research in Astronomy, Inc. under NASA contract NAS 5-26555. This work also uses data from MUSE, obtained from the ESO Science Archive Facility. 
RMcD is the recipient of an Australian Research Council Future Fellowship (project number FT150100333). 
The work of I.J. is supported by NOIRLab, which is managed by the Association of Universities for Research in Astronomy (AURA) under a cooperative agreement with the National Science Foundation.
The authors would also like to acknowledge Yunchong Wang for sharing simulation density slope data from  IllustrisTNG. 

%%%%%%%%%%%%%%%%%%%%%%%%%%%%%%%%%%%%%%%%%%%%%%%%%%
% data availability heading: 

\section*{Data Availability}
Values derived in this work are presented in Table \ref{full_data}. Sources of archival data from which this work is based are given in section \ref{datasample}.



\bibliographystyle{mnras}
\bibliography{good_ref} % if your bibtex file is called example.bib

\begin{thebibliography}{}
\makeatletter
\relax
\def\mn@urlcharsother{\let\do\@makeother \do\$\do\&\do\#\do\^\do\_\do\%\do\~}
\def\mn@doi{\begingroup\mn@urlcharsother \@ifnextchar [ {\mn@doi@}
  {\mn@doi@[]}}
\def\mn@doi@[#1]#2{\def\@tempa{#1}\ifx\@tempa\@empty \href
  {http://dx.doi.org/#2} {doi:#2}\else \href {http://dx.doi.org/#2} {#1}\fi
  \endgroup}
\def\mn@eprint#1#2{\mn@eprint@#1:#2::\@nil}
\def\mn@eprint@arXiv#1{\href {http://arxiv.org/abs/#1} {{\tt arXiv:#1}}}
\def\mn@eprint@dblp#1{\href {http://dblp.uni-trier.de/rec/bibtex/#1.xml}
  {dblp:#1}}
\def\mn@eprint@#1:#2:#3:#4\@nil{\def\@tempa {#1}\def\@tempb {#2}\def\@tempc
  {#3}\ifx \@tempc \@empty \let \@tempc \@tempb \let \@tempb \@tempa \fi \ifx
  \@tempb \@empty \def\@tempb {arXiv}\fi \@ifundefined
  {mn@eprint@\@tempb}{\@tempb:\@tempc}{\expandafter \expandafter \csname
  mn@eprint@\@tempb\endcsname \expandafter{\@tempc}}}

\bibitem[\protect\citeauthoryear{Auger, Treu, Bolton, Gavazzi, Koopmans,
  Marshall, Moustakas  \& Burles}{Auger et~al.}{2010}]{auger_sloan_2010}
Auger M.~W.,  Treu T.,  Bolton A.~S.,  Gavazzi R.,  Koopmans L. V.~E.,
  Marshall P.~J.,  Moustakas L.~A.,   Burles S.,  2010, \mn@doi [ApJ]
  {10.1088/0004-637X/724/1/511}, 724, 511

\bibitem[\protect\citeauthoryear{Bacon et~al.,}{Bacon
  et~al.}{2010}]{bacon_muse_2010}
Bacon R.,  et~al., 2010, \mn@doi [Proc. SPIE] {10.1117/12.856027}, 7735, 773508

\bibitem[\protect\citeauthoryear{Barbary}{Barbary}{2016}]{barbary_sep:_2016}
Barbary K.,  2016, \mn@doi [J. Open Source Softw.] {10.21105/joss.00058}, 1, 58

\bibitem[\protect\citeauthoryear{Barnabè, Czoske, Koopmans, Treu  \&
  Bolton}{Barnabè et~al.}{2011}]{barnabe_two-dimensional_2011}
Barnabè M.,  Czoske O.,  Koopmans L. V.~E.,  Treu T.,   Bolton A.~S.,  2011,
  \mn@doi [MNRAS] {10.1111/j.1365-2966.2011.18842.x}, 415, 2215

\bibitem[\protect\citeauthoryear{{Barnes} \& {Hernquist}}{{Barnes} \&
  {Hernquist}}{1991}]{barnes_1991_fueling}
{Barnes} J.~E.,  {Hernquist} L.~E.,  1991, \mn@doi [\apjl] {10.1086/185978},
  \href {https://ui.adsabs.harvard.edu/abs/1991ApJ...370L..65B} {370, L65}

\bibitem[\protect\citeauthoryear{Bellstedt et~al.,}{Bellstedt
  et~al.}{2018}]{bellstedt_sluggs_2018}
Bellstedt S.,  et~al., 2018, \mn@doi [MNRAS] {10.1093/mnras/sty456}, 476, 4543

\bibitem[\protect\citeauthoryear{Bertin \& Arnouts}{Bertin \&
  Arnouts}{1996}]{bertin_sextractor:_1996}
Bertin E.,  Arnouts S.,  1996, \mn@doi [A\&AS] {10.1051/aas:1996164}, 117, 393

\bibitem[\protect\citeauthoryear{Bezanson, van Dokkum, Tal, Marchesini, Kriek,
  Franx  \& Coppi}{Bezanson et~al.}{2009}]{bezanson_relation_2009}
Bezanson R.,  van Dokkum P.~G.,  Tal T.,  Marchesini D.,  Kriek M.,  Franx M.,
   Coppi P.,  2009, \mn@doi [ApJ] {10.1088/0004-637X/697/2/1290}, 697, 1290

\bibitem[\protect\citeauthoryear{Bezanson et~al.,}{Bezanson
  et~al.}{2018}]{bezanson_spatially_2018}
Bezanson R.,  et~al., 2018, \mn@doi [ApJ] {10.3847/1538-4357/aabc55}, 858, 60

\bibitem[\protect\citeauthoryear{Blumenthal, Faber, Primack  \&
  Rees}{Blumenthal et~al.}{1984}]{blumenthal_formation_1984}
Blumenthal G.~R.,  Faber S.~M.,  Primack J.~R.,   Rees M.~J.,  1984, \mn@doi
  [Nature] {10.1038/311517a0}, 311, 517

\bibitem[\protect\citeauthoryear{Bolton et~al.,}{Bolton
  et~al.}{2012}]{bolton_boss_2012}
Bolton A.~S.,  et~al., 2012, \mn@doi [ApJ] {10.1088/0004-637X/757/1/82}, 757,
  82

\bibitem[\protect\citeauthoryear{Brodie et~al.,}{Brodie
  et~al.}{2014}]{brodie_sages_2014}
Brodie J.~P.,  et~al., 2014, \mn@doi [ApJ] {10.1088/0004-637X/796/1/52}, 796,
  52

\bibitem[\protect\citeauthoryear{Brownstein et~al.,}{Brownstein
  et~al.}{2011}]{brownstein_boss_2011}
Brownstein J.~R.,  et~al., 2011, \mn@doi [ApJ] {10.1088/0004-637X/744/1/41},
  744, 41

\bibitem[\protect\citeauthoryear{Bundy, Fukugita, Ellis, Targett, Belli  \&
  Kodama}{Bundy et~al.}{2009}]{bundy_greater_2009}
Bundy K.,  Fukugita M.,  Ellis R.~S.,  Targett T.~A.,  Belli S.,   Kodama T.,
  2009, \mn@doi [ApJ] {10.1088/0004-637X/697/2/1369}, 697, 1369

\bibitem[\protect\citeauthoryear{Bundy et~al.,}{Bundy
  et~al.}{2014}]{bundy_overview_2014}
Bundy K.,  et~al., 2014, \mn@doi [ApJ] {10.1088/0004-637X/798/1/7}, 798, 7

\bibitem[\protect\citeauthoryear{Caminha et~al.,}{Caminha
  et~al.}{2017}]{caminha_refined_2017}
Caminha G.~B.,  et~al., 2017, \mn@doi [A&A] {10.1051/0004-6361/201629297}, 600,
  A90

\bibitem[\protect\citeauthoryear{Cappellari}{Cappellari}{2002}]{cappellari_efficient_2002}
Cappellari M.,  2002, \mn@doi [MNRAS] {10.1046/j.1365-8711.2002.05412.x}, 333,
  400

\bibitem[\protect\citeauthoryear{Cappellari}{Cappellari}{2008}]{cappellari_measuring_2008}
Cappellari M.,  2008, \mn@doi [MNRAS] {10.1111/j.1365-2966.2008.13754.x}, 390,
  71

\bibitem[\protect\citeauthoryear{Cappellari}{Cappellari}{2017}]{cappellari_improving_2017}
Cappellari M.,  2017, \mn@doi [MNRAS] {10.1093/mnras/stw3020}, 466, 798

\bibitem[\protect\citeauthoryear{Cappellari \& Copin}{Cappellari \&
  Copin}{2003}]{cappellari_adaptive_2003}
Cappellari M.,  Copin Y.,  2003, \mn@doi [MNRAS]
  {10.1046/j.1365-8711.2003.06541.x}, 342, 345

\bibitem[\protect\citeauthoryear{Cappellari \& Emsellem}{Cappellari \&
  Emsellem}{2004}]{cappellari_parametric_2004}
Cappellari M.,  Emsellem E.,  2004, \mn@doi [PASP] {10.1086/381875}, 116, 138

\bibitem[\protect\citeauthoryear{Cappellari et~al.,}{Cappellari
  et~al.}{2007}]{cappellari_sauron_2007}
Cappellari M.,  et~al., 2007, \mn@doi [MNRAS]
  {10.1111/j.1365-2966.2007.11963.x}, 379, 418

\bibitem[\protect\citeauthoryear{Cappellari et~al.,}{Cappellari
  et~al.}{2011}]{cappellari_atlas3d_2011}
Cappellari M.,  et~al., 2011, \mn@doi [MNRAS]
  {10.1111/j.1365-2966.2010.18174.x}, 413, 813

\bibitem[\protect\citeauthoryear{Cappellari et~al.,}{Cappellari
  et~al.}{2013}]{cappellari_atlas3d_2013-1}
Cappellari M.,  et~al., 2013, \mn@doi [MNRAS] {10.1093/mnras/stt562}, 432, 1709

\bibitem[\protect\citeauthoryear{Cappellari et~al.,}{Cappellari
  et~al.}{2015}]{cappellari_small_2015}
Cappellari M.,  et~al., 2015, \mn@doi [ApJ] {10.1088/2041-8205/804/1/L21}, 804,
  L21

\bibitem[\protect\citeauthoryear{Coccato et~al.,}{Coccato
  et~al.}{2009}]{coccato_kinematic_2009}
Coccato L.,  et~al., 2009, \mn@doi [MNRAS] {10.1111/j.1365-2966.2009.14417.x},
  394, 1249

\bibitem[\protect\citeauthoryear{Conselice}{Conselice}{2014}]{conselice_evolution_2014}
Conselice C.~J.,  2014, \mn@doi [Annu. Rev. A\&A]
  {10.1146/annurev-astro-081913-040037}, 52, 291

\bibitem[\protect\citeauthoryear{Dutton \& Treu}{Dutton \&
  Treu}{2014}]{dutton_bulgehalo_2014}
Dutton A.~A.,  Treu T.,  2014, \mn@doi [MNRAS] {10.1093/mnras/stt2489}, 438,
  3594

\bibitem[\protect\citeauthoryear{Díaz-García et~al.,}{Díaz-García
  et~al.}{2019}]{diaz-garcia_stellar_2019}
Díaz-García L.~A.,  et~al., 2019, \mn@doi [A/&A]
  {10.1051/0004-6361/201832882}, 631, A157

\bibitem[\protect\citeauthoryear{Eichner et~al.,}{Eichner
  et~al.}{2013}]{eichner_galaxy_2013}
Eichner T.,  et~al., 2013, \mn@doi [ApJ] {10.1088/0004-637X/774/2/124}, 774,
  124

\bibitem[\protect\citeauthoryear{Emsellem, Monnet  \& Bacon}{Emsellem
  et~al.}{1994}]{emsellem_multi-gaussian_1994}
Emsellem E.,  Monnet G.,   Bacon R.,  1994, A\&A, 285, 723

\bibitem[\protect\citeauthoryear{Falcón-Barroso, Sánchez-Blázquez, Vazdekis,
  Ricciardelli, Cardiel, Cenarro, Gorgas  \& Peletier}{Falcón-Barroso
  et~al.}{2011}]{falcon-barroso_updated_2011}
Falcón-Barroso J.,  Sánchez-Blázquez P.,  Vazdekis A.,  Ricciardelli E.,
  Cardiel N.,  Cenarro A.~J.,  Gorgas J.,   Peletier R.~F.,  2011, \mn@doi
  [A\&A] {10.1051/0004-6361/201116842}, 532, A95

\bibitem[\protect\citeauthoryear{Foreman-Mackey, Hogg, Lang  \&
  Goodman}{Foreman-Mackey et~al.}{2013}]{foreman-mackey_emcee:_2013}
Foreman-Mackey D.,  Hogg D.~W.,  Lang D.,   Goodman J.,  2013, \mn@doi [PASP]
  {10.1086/670067}, 125, 306

\bibitem[\protect\citeauthoryear{Foster et~al.,}{Foster
  et~al.}{2021}]{foster_2021_magpi}
Foster C.,  et~al., 2021 (\mn@eprint {arXiv} {2011.13567 (accepted)})

\bibitem[\protect\citeauthoryear{Franx, van Dokkum, Schreiber, Wuyts, Labbe  \&
  Toft}{Franx et~al.}{2008}]{franx_structure_2008}
Franx M.,  van Dokkum P.~G.,  Schreiber N. M.~F.,  Wuyts S.,  Labbe I.,   Toft
  S.,  2008, \mn@doi [ApJ] {10.1086/592431}, 688, 770

\bibitem[\protect\citeauthoryear{Gallazzi, Pasquali, Zibetti  \&
  Barbera}{Gallazzi et~al.}{2021}]{gallazzi_galaxy_2021}
Gallazzi A.~R.,  Pasquali A.,  Zibetti S.,   Barbera F.~L.,  2021, \mn@doi
  [MNRAS] {10.1093/mnras/stab265}, 502, 4457

\bibitem[\protect\citeauthoryear{Graaff et~al.,}{Graaff
  et~al.}{2020}]{graaff_tightly_2020}
Graaff A.~d.,  et~al., 2020, \mn@doi [ApJ] {10.3847/2041-8213/abc428}, 903, L30

\bibitem[\protect\citeauthoryear{Grillo et~al.,}{Grillo
  et~al.}{2015}]{grillo_clash-vlt_2015-1}
Grillo C.,  et~al., 2015, \mn@doi [ApJ] {10.1088/0004-637X/800/1/38}, 800, 38

\bibitem[\protect\citeauthoryear{Grillo et~al.,}{Grillo
  et~al.}{2016}]{grillo_story_2016}
Grillo C.,  et~al., 2016, \mn@doi [ApJ] {10.3847/0004-637X/822/2/78}, 822, 78

\bibitem[\protect\citeauthoryear{Hilz, Naab, Ostriker, Thomas, Burkert  \&
  Jesseit}{Hilz et~al.}{2012}]{hilz_relaxation_2012}
Hilz M.,  Naab T.,  Ostriker J.~P.,  Thomas J.,  Burkert A.,   Jesseit R.,
  2012, \mn@doi [MNRAS] {10.1111/j.1365-2966.2012.21541.x}, 425, 3119

\bibitem[\protect\citeauthoryear{Hilz, Naab  \& Ostriker}{Hilz
  et~al.}{2013}]{hilz_how_2013}
Hilz M.,  Naab T.,   Ostriker J.~P.,  2013, \mn@doi [MNRAS]
  {10.1093/mnras/sts501}, 429, 2924

\bibitem[\protect\citeauthoryear{Hogg \& Foreman-Mackey}{Hogg \&
  Foreman-Mackey}{2018}]{hogg_data_2018}
Hogg D.~W.,  Foreman-Mackey D.,  2018, \mn@doi [ApJS]
  {10.3847/1538-4365/aab76e}, 236, 11

\bibitem[\protect\citeauthoryear{Huertas-Company et~al.,}{Huertas-Company
  et~al.}{2013}]{huertas-company_evolution_2013}
Huertas-Company M.,  et~al., 2013, \mn@doi [MNRAS] {10.1093/mnras/sts150}, 428,
  1715

\bibitem[\protect\citeauthoryear{Humphrey \& Buote}{Humphrey \&
  Buote}{2010}]{humphrey_slope_2010}
Humphrey P.~J.,  Buote D.~A.,  2010, \mn@doi [MNRAS]
  {10.1111/j.1365-2966.2010.16257.x}, 403, 2143

\bibitem[\protect\citeauthoryear{Karademir, Remus, Burkert, Dolag, Hoffmann,
  Moster, Steinwandel  \& Zhang}{Karademir et~al.}{2019}]{karademir_outer_2019}
Karademir G.~S.,  Remus R.-S.,  Burkert A.,  Dolag K.,  Hoffmann T.~L.,  Moster
  B.~P.,  Steinwandel U.,   Zhang J.,  2019, \mn@doi [MNRAS]
  {10.1093/mnras/stz1251}, 487, 318

\bibitem[\protect\citeauthoryear{Karman et~al.,}{Karman
  et~al.}{2015}]{karman_muse_2015}
Karman W.,  et~al., 2015, \mn@doi [A\&A] {10.1051/0004-6361/201424962}, 574,
  A11

\bibitem[\protect\citeauthoryear{Khochfar \& Silk}{Khochfar \&
  Silk}{2006}]{khochfar_simple_2006}
Khochfar S.,  Silk J.,  2006, \mn@doi [ApJ] {10.1086/507768}, 648, L21

\bibitem[\protect\citeauthoryear{Koopmans \& Treu}{Koopmans \&
  Treu}{2004}]{koopmans_lenses_2004-1}
Koopmans L. V.~E.,  Treu T.,  2004, in Plionis M.,  ed., Multiwavelength
  {Cosmology}. Astrophysics and {Space} {Science} {Library}.
Springer Netherlands, Dordrecht, pp 23--26, \mn@doi{10.1007/0-306-48570-2_4}

\bibitem[\protect\citeauthoryear{Koopmans, Treu, Bolton, Burles  \&
  Moustakas}{Koopmans et~al.}{2006}]{koopmans_sloan_2006}
Koopmans L. V.~E.,  Treu T.,  Bolton A.~S.,  Burles S.,   Moustakas L.~A.,
  2006, \mn@doi [ApJ] {10.1086/505696}, 649, 599

\bibitem[\protect\citeauthoryear{Lagattuta et~al.,}{Lagattuta
  et~al.}{2017}]{lagattuta_lens_2017}
Lagattuta D.~J.,  et~al., 2017, \mn@doi [MNRAS] {10.1093/mnras/stx1079}, 469,
  3946

\bibitem[\protect\citeauthoryear{Larson et~al.,}{Larson
  et~al.}{2011}]{larson_seven-year_2011}
Larson D.,  et~al., 2011, \mn@doi [ApJS] {10.1088/0067-0049/192/2/16}, 192, 16

\bibitem[\protect\citeauthoryear{Lauer et~al.,}{Lauer
  et~al.}{1995}]{lauer_centers_1995}
Lauer T.~R.,  et~al., 1995, \mn@doi [AJ] {10.1086/117719}, 110, 2622

\bibitem[\protect\citeauthoryear{Li, Li, Mao, Xu, Long  \& Emsellem}{Li
  et~al.}{2016}]{li_assessing_2016}
Li H.,  Li R.,  Mao S.,  Xu D.,  Long R.~J.,   Emsellem E.,  2016, \mn@doi
  [MNRAS] {10.1093/mnras/stv2565}, 455, 3680

\bibitem[\protect\citeauthoryear{Li et~al.,}{Li et~al.}{2019}]{li_sdss-iv_2019}
Li R.,  et~al., 2019, \mn@doi [MNRAS] {10.1093/mnras/stz2565}, 490, 2124

\bibitem[\protect\citeauthoryear{Limousin, Kneib, Bardeau, Natarajan, Czoske,
  Smail, Ebeling  \& Smith}{Limousin et~al.}{2007}]{limousin_truncation_2007}
Limousin M.,  Kneib J.~P.,  Bardeau S.,  Natarajan P.,  Czoske O.,  Smail I.,
  Ebeling H.,   Smith G.~P.,  2007, \mn@doi [A\&A]
  {10.1051/0004-6361:20065543}, 461, 881

\bibitem[\protect\citeauthoryear{Lotz et~al.,}{Lotz
  et~al.}{2017}]{lotz_frontier_2017}
Lotz J.~M.,  et~al., 2017, \mn@doi [ApJ] {10.3847/1538-4357/837/1/97}, 837, 97

\bibitem[\protect\citeauthoryear{Lynden-Bell}{Lynden-Bell}{1967}]{lynden-bell_statistical_1967}
Lynden-Bell D.,  1967, \mn@doi [MNRAS] {10.1093/mnras/136.1.101}, 136, 101

\bibitem[\protect\citeauthoryear{Mahler et~al.,}{Mahler
  et~al.}{2018}]{mahler_strong_2018}
Mahler G.,  et~al., 2018, \mn@doi [MNRAS] {10.1093/mnras/stx1971}, 473, 663

\bibitem[\protect\citeauthoryear{Mandelbaum, van~de Ven  \& Keeton}{Mandelbaum
  et~al.}{2009}]{mandelbaum_galaxy_2009}
Mandelbaum R.,  van~de Ven G.,   Keeton C.~R.,  2009, \mn@doi [MNRAS]
  {10.1111/j.1365-2966.2009.15166.x}, 398, 635

\bibitem[\protect\citeauthoryear{McDermid et~al.,}{McDermid
  et~al.}{2015}]{mcdermid_atlas3d_2015}
McDermid R.~M.,  et~al., 2015, \mn@doi [MNRAS] {10.1093/mnras/stv105}, 448,
  3484

\bibitem[\protect\citeauthoryear{McDermid et~al.,}{McDermid
  et~al.}{2020}]{mcdermid_phase_2020}
McDermid R.~M.,  et~al., 2020 (\mn@eprint {arXiv} {2009.09242})

\bibitem[\protect\citeauthoryear{McLure et~al.,}{McLure
  et~al.}{2013}]{mclure_sizes_2013}
McLure R.~J.,  et~al., 2013, \mn@doi [MNRAS] {10.1093/mnras/sts092}, 428, 1088

\bibitem[\protect\citeauthoryear{Merten et~al.,}{Merten
  et~al.}{2011}]{merten_creation_2011}
Merten J.,  et~al., 2011, \mn@doi [MNRAS] {10.1111/j.1365-2966.2011.19266.x},
  417, 333

\bibitem[\protect\citeauthoryear{Meylan, Jetzer, North, Schneider, Kochanek  \&
  Wambsganss}{Meylan et~al.}{2006}]{meylan_gravitational_2006}
Meylan G.,  Jetzer P.,  North P.,  Schneider P.,  Kochanek C.~S.,   Wambsganss
  J.,  2006, Saas-Fee Advanced Course 33: Gravitational Lensing: Strong, Weak
  and Micro

\bibitem[\protect\citeauthoryear{Mihos \& Hernquist}{Mihos \&
  Hernquist}{1994}]{mihos_ultraluminous_1994}
Mihos J.~C.,  Hernquist L.,  1994, \mn@doi [ApJ] {10.1086/187460}, 431, L9

\bibitem[\protect\citeauthoryear{Monnet, Bacon  \& Emsellem}{Monnet
  et~al.}{1992}]{monnet_modelling_1992}
Monnet G.,  Bacon R.,   Emsellem E.,  1992, A\&A, 253, 366

\bibitem[\protect\citeauthoryear{Mowla et~al.,}{Mowla
  et~al.}{2019}]{mowla_cosmos-dash_2019}
Mowla L.,  et~al., 2019, \mn@doi [ApJ] {10.3847/1538-4357/ab290a}, 880, 57

\bibitem[\protect\citeauthoryear{Naab, Johansson  \& Ostriker}{Naab
  et~al.}{2009}]{naab_minor_2009}
Naab T.,  Johansson P.~H.,   Ostriker J.~P.,  2009, \mn@doi [ApJ]
  {10.1088/0004-637X/699/2/L178}, 699, L178

\bibitem[\protect\citeauthoryear{Navarro, Frenk  \& White}{Navarro
  et~al.}{1997}]{navarro_universal_1997}
Navarro J.~F.,  Frenk C.~S.,   White S. D.~M.,  1997, \mn@doi [ApJ]
  {10.1086/304888}, 490, 493

\bibitem[\protect\citeauthoryear{Newman, Ellis, Bundy  \& Treu}{Newman
  et~al.}{2012}]{newman_can_2012}
Newman A.~B.,  Ellis R.~S.,  Bundy K.,   Treu T.,  2012, \mn@doi [ApJ]
  {10.1088/0004-637X/746/2/162}, 746, 162

\bibitem[\protect\citeauthoryear{Nipoti, Treu  \& Bolton}{Nipoti
  et~al.}{2009}]{nipoti_dry_2009-1}
Nipoti C.,  Treu T.,   Bolton A.~S.,  2009, \mn@doi [ApJ]
  {10.1088/0004-637X/703/2/1531}, 703, 1531

\bibitem[\protect\citeauthoryear{Oogi \& Habe}{Oogi \&
  Habe}{2013}]{oogi_dry_2013}
Oogi T.,  Habe A.,  2013, \mn@doi [MNRAS] {10.1093/mnras/sts047}, 428, 641

\bibitem[\protect\citeauthoryear{Oser, Ostriker, Naab, Johansson  \&
  Burkert}{Oser et~al.}{2010}]{oser_two_2010}
Oser L.,  Ostriker J.~P.,  Naab T.,  Johansson P.~H.,   Burkert A.,  2010,
  \mn@doi [ApJ] {10.1088/0004-637X/725/2/2312}, 725, 2312

\bibitem[\protect\citeauthoryear{Pillepich et~al.,}{Pillepich
  et~al.}{2018}]{pillepich_simulating_2018}
Pillepich A.,  et~al., 2018, \mn@doi [MNRAS] {10.1093/mnras/stx2656}, 473, 4077

\bibitem[\protect\citeauthoryear{Poci, Cappellari  \& McDermid}{Poci
  et~al.}{2017}]{poci_systematic_2017}
Poci A.,  Cappellari M.,   McDermid R.~M.,  2017, \mn@doi [MNRAS]
  {10.1093/mnras/stx101}, p. stx101

\bibitem[\protect\citeauthoryear{Poggianti, Moretti, Calvi, D'Onofrio,
  Valentinuzzi, Fritz  \& Renzini}{Poggianti
  et~al.}{2013}]{poggianti_evolution_2013-1}
Poggianti B.~M.,  Moretti A.,  Calvi R.,  D'Onofrio M.,  Valentinuzzi T.,
  Fritz J.,   Renzini A.,  2013, \mn@doi [ApJ] {10.1088/0004-637X/777/2/125},
  777, 125

\bibitem[\protect\citeauthoryear{Remus, Burkert, Dolag, Johansson, Naab, Oser
  \& Thomas}{Remus et~al.}{2013}]{remus_dark_2013}
Remus R.-S.,  Burkert A.,  Dolag K.,  Johansson P.~H.,  Naab T.,  Oser L.,
  Thomas J.,  2013, \mn@doi [ApJ] {10.1088/0004-637X/766/2/71}, 766, 71

\bibitem[\protect\citeauthoryear{Remus, Dolag, Naab, Burkert, Hirschmann,
  Hoffmann  \& Johansson}{Remus et~al.}{2017}]{remus_co-evolution_2017}
Remus R.-S.,  Dolag K.,  Naab T.,  Burkert A.,  Hirschmann M.,  Hoffmann T.~L.,
    Johansson P.~H.,  2017, \mn@doi [MNRAS] {10.1093/mnras/stw2594}, 464, 3742

\bibitem[\protect\citeauthoryear{{Renaud}, {Theis}  \& {Boily}}{{Renaud}
  et~al.}{2008}]{renaud_2008_starbust}
{Renaud} F.,  {Theis} C.,   {Boily} C.~M.,  2008, \mn@doi [Astro. Nachr]
  {10.1002/asna.200811083}, \href
  {https://ui.adsabs.harvard.edu/abs/2008AN....329.1050R} {329, 1050}

\bibitem[\protect\citeauthoryear{Richard, Kneib, Limousin, Edge  \&
  Jullo}{Richard et~al.}{2010}]{richard_abell_2010}
Richard J.,  Kneib J.-P.,  Limousin M.,  Edge A.,   Jullo E.,  2010, \mn@doi
  [MNRAS] {10.1111/j.1745-3933.2009.00796.x}, 402, L44

\bibitem[\protect\citeauthoryear{Robertson, Hernquist, Cox, Di~Matteo, Hopkins,
  Martini  \& Springel}{Robertson et~al.}{2006}]{robertson_evolution_2006}
Robertson B.,  Hernquist L.,  Cox T.~J.,  Di~Matteo T.,  Hopkins P.~F.,
  Martini P.,   Springel V.,  2006, \mn@doi [ApJ] {10.1086/500348}, 641, 90

\bibitem[\protect\citeauthoryear{Ruff, Gavazzi, Marshall, Treu, Auger  \&
  Brault}{Ruff et~al.}{2011}]{ruff_sl2s_2011}
Ruff A.~J.,  Gavazzi R.,  Marshall P.~J.,  Treu T.,  Auger M.~W.,   Brault F.,
  2011, \mn@doi [ApJ] {10.1088/0004-637X/727/2/96}, 727, 96

\bibitem[\protect\citeauthoryear{Salpeter}{Salpeter}{1955}]{salpeter_luminosity_1955}
Salpeter E.~E.,  1955, \mn@doi [ApJ] {10.1086/145971}, 121, 161

\bibitem[\protect\citeauthoryear{{Sanders} \& {Mirabel}}{{Sanders} \&
  {Mirabel}}{1996}]{sanders_1996_luminous}
{Sanders} D.~B.,  {Mirabel} I.~F.,  1996, \mn@doi [\araa]
  {10.1146/annurev.astro.34.1.749}, \href
  {https://ui.adsabs.harvard.edu/abs/1996ARA&A..34..749S} {34, 749}

\bibitem[\protect\citeauthoryear{Scott et~al.,}{Scott
  et~al.}{2013}]{scott_atlas3d_2013-1}
Scott N.,  et~al., 2013, \mn@doi [MNRAS] {10.1093/mnras/sts422}, 432, 1894

\bibitem[\protect\citeauthoryear{Serra, Oosterloo, Cappellari, Heijer  \&
  Józsa}{Serra et~al.}{2016}]{serra_linear_2016}
Serra P.,  Oosterloo T.,  Cappellari M.,  Heijer M.~d.,   Józsa G. I.~G.,
  2016, \mn@doi [MNRAS] {10.1093/mnras/stw1010}, 460, 1382

\bibitem[\protect\citeauthoryear{Shajib, Treu, Birrer  \& Sonnenfeld}{Shajib
  et~al.}{2021}]{shajib_dark_2021}
Shajib A.~J.,  Treu T.,  Birrer S.,   Sonnenfeld A.,  2021, \mn@doi [MNRAS]
  {10.1093/mnras/stab536}, 503, 2380

\bibitem[\protect\citeauthoryear{Shankar, Marulli, Bernardi, Mei, Meert  \&
  Vikram}{Shankar et~al.}{2013}]{shankar_size_2013}
Shankar F.,  Marulli F.,  Bernardi M.,  Mei S.,  Meert A.,   Vikram V.,  2013,
  \mn@doi [MNRAS] {10.1093/mnras/sts001}, 428, 109

\bibitem[\protect\citeauthoryear{Sonnenfeld, Treu, Gavazzi, Suyu, Marshall,
  Auger  \& Nipoti}{Sonnenfeld et~al.}{2013}]{sonnenfeld_sl2s_2013}
Sonnenfeld A.,  Treu T.,  Gavazzi R.,  Suyu S.~H.,  Marshall P.~J.,  Auger
  M.~W.,   Nipoti C.,  2013, \mn@doi [ApJ] {10.1088/0004-637X/777/2/98}, 777,
  98

\bibitem[\protect\citeauthoryear{Sonnenfeld, Nipoti  \& Treu}{Sonnenfeld
  et~al.}{2014}]{sonnenfeld_purely_2014}
Sonnenfeld A.,  Nipoti C.,   Treu T.,  2014, \mn@doi [ApJ]
  {10.1088/0004-637X/786/2/89}, 786, 89

\bibitem[\protect\citeauthoryear{Springel et~al.,}{Springel
  et~al.}{2018}]{springel_first_2018}
Springel V.,  et~al., 2018, \mn@doi [MNRAS] {10.1093/mnras/stx3304}, 475, 676

\bibitem[\protect\citeauthoryear{Teklu, Remus, Dolag, Beck, Burkert, Schmidt,
  Schulze  \& Steinborn}{Teklu et~al.}{2015}]{teklu_connecting_2015}
Teklu A.~F.,  Remus R.-S.,  Dolag K.,  Beck A.~M.,  Burkert A.,  Schmidt A.~S.,
   Schulze F.,   Steinborn L.~K.,  2015, \mn@doi [ApJ]
  {10.1088/0004-637X/812/1/29}, 812, 29

\bibitem[\protect\citeauthoryear{Thomas, Maraston, Bender  \& Mendes~de
  Oliveira}{Thomas et~al.}{2005}]{thomas_epochs_2005}
Thomas D.,  Maraston C.,  Bender R.,   Mendes~de Oliveira C.,  2005, \mn@doi
  [ApJ] {10.1086/426932}, 621, 673

\bibitem[\protect\citeauthoryear{Tortorelli et~al.,}{Tortorelli
  et~al.}{2018}]{tortorelli_kormendy_2018}
Tortorelli L.,  et~al., 2018, \mn@doi [MNRAS] {10.1093/mnras/sty617}, 477, 648

\bibitem[\protect\citeauthoryear{Trager, Faber, Worthey  \& Gonzalez}{Trager
  et~al.}{2000}]{trager_stellar_2000}
Trager S.~C.,  Faber S.~M.,  Worthey G.,   Gonzalez J.~J.,  2000, \mn@doi [AJ]
  {10.1086/301442}, 120, 165

\bibitem[\protect\citeauthoryear{Treu \& Koopmans}{Treu \&
  Koopmans}{2004}]{treu_massive_2004}
Treu T.,  Koopmans L. V.~E.,  2004, \mn@doi [ApJ] {10.1086/422245}, 611, 739

\bibitem[\protect\citeauthoryear{Trujillo, Conselice, Bundy, Cooper, Eisenhardt
   \& Ellis}{Trujillo et~al.}{2007}]{trujillo_strong_2007}
Trujillo I.,  Conselice C.~J.,  Bundy K.,  Cooper M.~C.,  Eisenhardt P.,
  Ellis R.~S.,  2007, \mn@doi [MNRAS] {10.1111/j.1365-2966.2007.12388.x}, 382,
  109

\bibitem[\protect\citeauthoryear{\VAN{Dokkum}{Van}{van}~Dokkum
  et~al.,}{\VAN{Dokkum}{Van}{van}~Dokkum et~al.}{2010}]{van_dokkum_growth_2010}
\VAN{Dokkum}{Van}{van}~Dokkum P.~G.,  et~al., 2010, \mn@doi [\apj]
  {10.1088/0004-637X/709/2/1018}, \href
  {https://ui.adsabs.harvard.edu/abs/2010ApJ...709.1018V} {709, 1018}

\bibitem[\protect\citeauthoryear{\VAN{Sande}{Van de}{van de} Sande
  et~al.,}{\VAN{Sande}{Van de}{van de} et~al.}{2017}]{van_de_sande_sami_2017}
\VAN{Sande}{Van de}{van de} Sande J.,  et~al., 2017, \mn@doi [ApJ]
  {10.3847/1538-4357/835/1/104}, 835, 104

\bibitem[\protect\citeauthoryear{\VAN{Wel}{Van}{van der}~Wel
  et~al.,}{\VAN{Wel}{Van}{van der}~Wel
  et~al.}{2014}]{van_der_wel_3d-hstcandels_2014}
\VAN{Wel}{Van}{van der}~Wel A.,  et~al., 2014, \mn@doi [ApJ]
  {10.1088/0004-637X/788/1/28}, 788, 28

\bibitem[\protect\citeauthoryear{Vaughan, Davies, Zieleniewski  \&
  Houghton}{Vaughan et~al.}{2018}]{vaughan_stellar_2018}
Vaughan S.~P.,  Davies R.~L.,  Zieleniewski S.,   Houghton R. C.~W.,  2018,
  \mn@doi [MNRAS] {10.1093/mnras/sty1434}, 479, 2443

\bibitem[\protect\citeauthoryear{Vazdekis, Sánchez-Blázquez, Falcón-Barroso,
  Cenarro, Beasley, Cardiel, Gorgas  \& Peletier}{Vazdekis
  et~al.}{2010}]{vazdekis_evolutionary_2010}
Vazdekis A.,  Sánchez-Blázquez P.,  Falcón-Barroso J.,  Cenarro A.~J.,
  Beasley M.~A.,  Cardiel N.,  Gorgas J.,   Peletier R.~F.,  2010, \mn@doi
  [MNRAS] {10.1111/j.1365-2966.2010.16407.x}

\bibitem[\protect\citeauthoryear{Wang et~al.,}{Wang
  et~al.}{2019}]{wang_early-type_2019-1}
Wang Y.,  et~al., 2019, \mn@doi [MNRAS] {10.1093/mnras/stz2907}, 490, 5722

\bibitem[\protect\citeauthoryear{Wang et~al.,}{Wang
  et~al.}{2020}]{wang_early-type_2020}
Wang Y.,  et~al., 2020, \mn@doi [MNRAS] {10.1093/mnras/stz3348}, 491, 5188

\bibitem[\protect\citeauthoryear{Weijmans, Krajnović, van~de Ven, Oosterloo,
  Morganti  \& de Zeeuw}{Weijmans et~al.}{2008}]{weijmans_shape_2008}
Weijmans A.-M.,  Krajnović D.,  van~de Ven G.,  Oosterloo T.~A.,  Morganti R.,
    de Zeeuw P.~T.,  2008, \mn@doi [MNRAS] {10.1111/j.1365-2966.2007.12680.x},
  383, 1343

\bibitem[\protect\citeauthoryear{Weilbacher et~al.,}{Weilbacher
  et~al.}{2020}]{weilbacher_data_2020}
Weilbacher P.~M.,  et~al., 2020, \mn@doi [A\&A] {10.1051/0004-6361/202037855},
  641, A28

\bibitem[\protect\citeauthoryear{White \& Rees}{White \&
  Rees}{1978}]{white_core_1978}
White S. D.~M.,  Rees M.~J.,  1978, \mn@doi [MNRAS] {10.1093/mnras/183.3.341},
  183, 341

\bibitem[\protect\citeauthoryear{{Whitmore} \& {Schweizer}}{{Whitmore} \&
  {Schweizer}}{1995}]{whitmore_1995_hubble}
{Whitmore} B.~C.,  {Schweizer} F.,  1995, \mn@doi [\aj] {10.1086/117334}, \href
  {https://ui.adsabs.harvard.edu/abs/1995AJ....109..960W} {109, 960}

\bibitem[\protect\citeauthoryear{Williamson et~al.,}{Williamson
  et~al.}{2011}]{williamson_sunyaev-zeltextquotesingledovich-selected_2011}
Williamson R.,  et~al., 2011, \mn@doi [ApJ] {10.1088/0004-637X/738/2/139}, 738,
  139

\bibitem[\protect\citeauthoryear{Wright}{Wright}{2006}]{wright_cosmology_2006}
Wright E.~L.,  2006, \mn@doi [PASP] {10.1086/510102}, 118, 1711

\bibitem[\protect\citeauthoryear{Xu, Springel, Sluse, Schneider, Sonnenfeld,
  Nelson, Vogelsberger  \& Hernquist}{Xu et~al.}{2017}]{xu_inner_2017}
Xu D.,  Springel V.,  Sluse D.,  Schneider P.,  Sonnenfeld A.,  Nelson D.,
  Vogelsberger M.,   Hernquist L.,  2017, \mn@doi [MNRAS]
  {10.1093/mnras/stx899}, 469, 1824

\bibitem[\protect\citeauthoryear{Zheng et~al.,}{Zheng
  et~al.}{2012}]{zheng_magnified_2012}
Zheng W.,  et~al., 2012, \mn@doi [Nature] {10.1038/nature11446}, 489, 406

\makeatother
\end{thebibliography}

%%%%%%%%%%%%%%%%%%%%%%%%%%%%%%%%%%%%%%%%%%%%%%%%%%

%%%%%%%%%%%%%%%%% APPENDICES %%%%%%%%%%%%%%%%%%%%%

\appendix
\section{Galaxy Parameters}
Figure \ref{appendixfigure} shows the HST MGEs and the MUSE stellar kinematic maps for all 90 galaxies in the sample. The \textsc{emcee} posterior distribution for the inner density slope $\gamma'$ is also shown, along with the modelled $v_\text{rms}$ fields. Table \ref{full_data} provides derived values for the galaxies studied in this work, including coordinates, redshifts, stellar density slopes, and total density slopes.

% result table: 
% cluster, ID, sigma_e, R_e(kpc), density_slope, redshift. 
\begin{table*}
\centering
\caption{Derived parameters for the 90 galaxies used in this work. Cluster names are given in  redshift order (column 1) with IDs as defined in Section \ref{datasample} (column 2) and galaxy coordinates given in column 3. Redshifts as found by pPXF using an circular aperture spectrum with the associated aperture velocity dispersions are given in columns 4 and 5. The circularised effective radius of each galaxy is given in column 6, based on the MGE fit. Derived dynamical masses are given in column 7, computed within an effective radius. Stellar density slopes are given in column 8. The stellar density slopes have no associated formal error as they are derived directly from HST photometry MGEs, and systematic effects, such as the radial limits used, dominate. The total mass density slopes are given in column 9, with the errors corresponding to the 16th and 84th percentiles of the {\sc{emcee}} distribution.}
\label{full_data}
    \begin{tabular}{lcccccccc} \hline
      Cluster & ID & Coordinates (J2000) & $z$ & $\sigma_{\text{e}} \,\,(\text{kms}^{-1})$  & $\text{R}_{\text{e}}$ (kpc) & $\log_{10}(\text{M}/\text{M}_{\odot})$ &$\gamma_{\star}$& $\gamma$\\\hline
      (1) & (2) & (3) & (4) & (5) &  (6) & (7) & (8) & (9) \\\hline 
      A2744 & 2284 & 3.60265,\,-30.41696 & 0.3128 & 123 & 1.89 & 10.56 & -2.71 & $-2.11_{-0.10}^{+0.11}$ \\
A2744 & 3540 & 3.58882,\,-30.41072 & 0.3218 & 110 & 1.73 & 10.35 & -2.76 & $-1.39_{-0.19}^{+0.21}$ \\
A2744 & 3699 & 3.58251,\,-30.40999 & 0.3184 & 130 & 0.93 & 10.27 & -3.23 & $-1.47_{-0.25}^{+0.26}$ \\
A2744 & 3870 & 3.59512,\,-30.40937 & 0.3195 & 164 & 1.43 & 10.66 & -3.84 & $-2.32_{-0.28}^{+0.28}$ \\
A2744 & 3910 & 3.58913,\,-30.40957 & 0.3171 & 109 & 1.32 & 10.26 & -2.67 & $-1.99_{-0.13}^{+0.13}$ \\
A2744 & 4423 & 3.57967,\,-30.40919 & 0.3022 & 188 & 2.64 & 10.99 & -2.63 & $-2.25_{-0.02}^{+0.02}$ \\
A2744 & 4439 & 3.59028,\,-30.40740 & 0.3178 & 108 & 2.77 & 10.57 & -4.40 & $-2.20_{-0.27}^{+0.26}$ \\
A2744 & 4556 & 3.59172,\,-30.40781 & 0.3189 & 195 & 1.42 & 10.80 & -3.19 & $-2.05_{-0.11}^{+0.11}$ \\
A2744 & 5061 & 3.57394,\,-30.40883 & 0.3131 & 183 & 1.97 & 10.88 & -2.67 & $-2.11_{-0.05}^{+0.06}$ \\
A2744 & 5339 & 3.59591,\,-30.40621 & 0.3156 & 143 & 2.38 & 10.75 & -2.86 & $-1.89_{-0.09}^{+0.09}$ \\
A2744 & 5693 & 3.58704,\,-30.40495 & 0.2983 & 157 & 1.45 & 10.67 & -3.59 & $-1.84_{-0.11}^{+0.12}$ \\
A2744 & 6043 & 3.58437,\,-30.40289 & 0.3155 & 87 & 2.10 & 10.32 & -3.14 & $-1.93_{-0.32}^{+0.32}$ \\
A2744 & 7068 & 3.60527,\,-30.40081 & 0.3192 & 248 & 0.71 & 10.73 & -3.46 & $-2.36_{-0.12}^{+0.14}$ \\
A2744 & 7229 & 3.59446,\,-30.40035 & 0.3031 & 174 & 0.79 & 10.44 & -3.47 & $-1.98_{-0.18}^{+0.17}$ \\
A2744 & 7344 & 3.60435,\,-30.40013 & 0.3186 & 163 & 1.84 & 10.74 & -2.65 & $-2.09_{-0.06}^{+0.07}$ \\
A2744 & 7947 & 3.59300,\,-30.39933 & 0.3087 & 149 & 1.48 & 10.58 & -2.60 & $-2.12_{-0.02}^{+0.02}$ \\
A2744 & 8067 & 3.57491,\,-30.39838 & 0.3171 & 185 & 2.52 & 10.98 & -2.46 & $-2.21_{-0.07}^{+0.06}$ \\
A2744 & 8117 & 3.58216,\,-30.39857 & 0.2981 & 168 & 1.93 & 10.84 & -2.94 & $-2.07_{-0.06}^{+0.06}$ \\
A2744 & 8729 & 3.57886,\,-30.39712 & 0.3186 & 123 & 2.10 & 10.56 & -2.44 & $-2.01_{-0.12}^{+0.14}$ \\
A2744 & 9072 & 3.59896,\,-30.39752 & 0.3157 & 101 & 4.98 & 10.67 & -2.26 & $-2.17_{-0.07}^{+0.07}$ \\
A2744 & 9283 & 3.60083,\,-30.39490 & 0.3058 & 92 & 1.57 & 10.20 & -2.77 & $-2.11_{-0.28}^{+0.26}$ \\
A2744 & 9646 & 3.57895,\,-30.39412 & 0.3187 & 135 & 2.19 & 10.68 & -2.78 & $-2.09_{-0.11}^{+0.11}$ \\
A2744 & 9710 & 3.58498,\,-30.39287 & 0.2951 & 98 & 1.11 & 10.03 & -3.01 & $-1.71_{-0.13}^{+0.12}$ \\
A2744 & 9876 & 3.58037,\,-30.39220 & 0.2933 & 81 & 1.24 & 9.97 & -2.77 & $-2.00_{-0.21}^{+0.20}$ \\
A2744 & 10314 & 3.59034,\,-30.39094 & 0.2967 & 150 & 1.25 & 10.53 & -3.15 & $-2.31_{-0.07}^{+0.07}$ \\
A2744 & 10478 & 3.57150,\,-30.39043 & 0.2960 & 96 & 2.63 & 10.36 & -2.57 & $-2.38_{-0.11}^{+0.10}$ \\
A2744 & 10689 & 3.59480,\,-30.39165 & 0.3002 & 218 & 4.32 & 11.38 & -2.35 & $-1.90_{-0.02}^{+0.02}$ \\
A2744 & 10884 & 3.59028,\,-30.38269 & 0.3014 & 123 & 1.31 & 10.37 & -2.86 & $-1.87_{-0.16}^{+0.14}$ \\
A2744 & 11363 & 3.58815,\,-30.38500 & 0.2972 & 149 & 1.27 & 10.50 & -2.78 & $-2.15_{-0.08}^{+0.07}$ \\
A2744 & 11418 & 3.59255,\,-30.38531 & 0.3160 & 200 & 0.84 & 10.59 & -3.35 & $-2.25_{-0.17}^{+0.17}$ \\
A2744 & 11440 & 3.60543,\,-30.38484 & 0.3108 & 180 & 1.15 & 10.68 & -3.00 & $-2.19_{-0.06}^{+0.06}$ \\
A2744 & 11856 & 3.58531,\,-30.38755 & 0.3002 & 107 & 1.72 & 10.39 & -2.57 & $-2.31_{-0.05}^{+0.04}$ \\
A2744 & 11950 & 3.58919,\,-30.38740 & 0.3164 & 195 & 1.30 & 10.79 & -2.98 & $-2.32_{-0.06}^{+0.06}$ \\
A2744 & 12149 & 3.59877,\,-30.38802 & 0.3022 & 122 & 1.12 & 10.33 & -3.10 & $-1.90_{-0.14}^{+0.13}$ \\
A2744 & 12269 & 3.59551,\,-30.38868 & 0.3027 & 112 & 1.48 & 10.33 & -2.91 & $-2.11_{-0.19}^{+0.18}$ \\
A2744 & 12443 & 3.59471,\,-30.38912 & 0.3030 & 95 & 3.18 & 10.46 & -2.34 & $-2.27_{-0.10}^{+0.10}$ \\
As1063 & 19 & 342.17703,\,-44.53694 & 0.3366 & 136 & 2.84 & 10.84 & -2.62 & $-1.77_{-0.12}^{+0.13}$ \\
As1063 & 25 & 342.17903,\,-44.53278 & 0.3478 & 292 & 2.63 & 11.41 & -3.13 & $-1.76_{-0.27}^{+0.25}$ \\
As1063 & 39 & 342.18406,\,-44.52692 & 0.3505 & 133 & 2.53 & 10.77 & -2.91 & $-1.77_{-0.20}^{+0.20}$ \\
As1063 & 41 & 342.18438,\,-44.53619 & 0.3535 & 157 & 2.78 & 10.91 & -2.63 & $-2.22_{-0.41}^{+0.41}$ \\
As1063 & 45 & 342.18542,\,-44.51863 & 0.3433 & 263 & 1.30 & 11.04 & -3.03 & $-2.17_{-0.08}^{+0.07}$ \\
As1063 & 51 & 342.18665,\,-44.52247 & 0.3386 & 140 & 1.07 & 10.43 & -3.21 & $-2.28_{-0.51}^{+0.35}$ \\
As1063 & 58 & 342.18813,\,-44.52595 & 0.3492 & 306 & 1.13 & 11.14 & -3.18 & $-2.33_{-0.11}^{+0.13}$ \\
\hline
\end{tabular}
\end{table*}

\begin{table*}
\centering
\contcaption{}
%\caption{Key properties of the chosen galaxy clusters for this project. Scales at each redshift using the cosmology calculator of \citet{wright_cosmology_2006}. The exposure time is the total for stacked fields, therefore each individual galaxy can have a lower exposure time.}
%\label{full_data}
    \begin{tabular}{lcccccccc} \hline
      Cluster & ID & Coordinates (J2000)& $z$ & $\sigma_{\text{e}} \,\,(\text{kms}^{-1})$  & $\text{R}_{\text{e}}$ (kpc) & $\log_{10}(\text{M}/\text{M}_{\odot})$ &$\gamma_{\star}$ & $\gamma$\\\hline
      (1) & (2) & (3) & (4) & (5) &  (6) & (7) & (8) & 9  \\\hline
      As1063 & 59 & 342.18814,\,-44.52972 & 0.3492 & 198 & 2.25 & 11.02 & -3.00 & $-1.71_{-0.10}^{+0.10}$ \\
As1063 & 78 & 342.19330,\,-44.51782 & 0.3418 & 314 & 0.96 & 11.12 & -3.28 & $-2.46_{-0.11}^{+0.12}$ \\
As1063 & 79 & 342.19330,\,-44.52643 & 0.3445 & 182 & 1.39 & 10.72 & -3.22 & $-1.90_{-0.17}^{+0.18}$ \\
As1063 & 85 & 342.19552,\,-44.52599 & 0.3455 & 245 & 2.81 & 11.32 & -2.87 & $-2.29_{-0.04}^{+0.04}$ \\
As1063 & 87 & 342.19727,\,-44.52327 & 0.3507 & 174 & 1.37 & 10.72 & -3.24 & $-2.05_{-0.33}^{+0.31}$ \\
As1063 & 90 & 342.20029,\,-44.52520 & 0.3524 & 138 & 2.15 & 10.64 & -2.59 & $-1.86_{-0.25}^{+0.22}$ \\
As1063 & 94 & 342.20418,\,-44.52524 & 0.3502 & 139 & 1.41 & 10.53 & -2.63 & $-2.23_{-0.23}^{+0.24}$ \\
A370 & Cl6 & 39.96772,\,-1.58660 & 0.3637 & 178 & 1.87 & 10.88 & -2.59 & $-2.05_{-0.12}^{+0.10}$ \\
A370 & Cl12 & 39.97745,\,-1.57645 & 0.3676 & 180 & 2.40 & 10.97 & -2.53 & $-2.12_{-0.10}^{+0.10}$ \\
A370 & Cl16 & 39.97181,\,-1.57478 & 0.3681 & 176 & 1.66 & 10.74 & -2.85 & $-2.05_{-0.25}^{+0.27}$ \\
A370 & Cl18 & 39.96540,\,-1.58602 & 0.3699 & 212 & 1.51 & 10.89 & -2.83 & $-2.16_{-0.09}^{+0.10}$ \\
A370 & Cl19 & 39.96960,\,-1.58380 & 0.3702 & 195 & 3.74 & 11.22 & -2.62 & $-2.12_{-0.04}^{+0.04}$ \\
A370 & Cl20 & 39.97050,\,-1.57488 & 0.3701 & 113 & 3.28 & 10.66 & -2.77 & $-1.97_{-0.29}^{+0.30}$ \\
A370 & Cl21 & 39.96378,\,-1.58104 & 0.3709 & 244 & 3.89 & 11.40 & -2.24 & $-2.30_{-0.03}^{+0.03}$ \\
A370 & Cl24 & 39.97726,\,-1.58191 & 0.3710 & 293 & 1.38 & 11.12 & -3.00 & $-2.37_{-0.10}^{+0.09}$ \\
A370 & Cl31 & 39.96840,\,-1.57469 & 0.3740 & 192 & 3.29 & 11.09 & -2.58 & $-2.16_{-0.07}^{+0.07}$ \\
A370 & Cl33 & 39.97112,\,-1.58690 & 0.3747 & 204 & 1.81 & 10.94 & -2.89 & $-2.03_{-0.13}^{+0.14}$ \\
A370 & Cl35 & 39.97060,\,-1.58378 & 0.3754 & 240 & 6.76 & 11.77 & -3.01 & $-1.75_{-0.13}^{+0.14}$ \\
A370 & Cl37 & 39.97514,\,-1.57687 & 0.3760 & 137 & 1.57 & 10.59 & -3.00 & $-2.12_{-0.13}^{+0.13}$ \\
A370 & Cl38 & 39.96807,\,-1.57563 & 0.3779 & 157 & 2.99 & 10.96 & -3.02 & $-2.09_{-0.22}^{+0.26}$ \\
A370 & Cl46 & 39.97579,\,-1.58581 & 0.3806 & 288 & 2.10 & 11.31 & -2.92 & $-2.21_{-0.04}^{+0.04}$ \\
A370 & Cl55 & 39.96909,\,-1.57869 & 0.3882 & 275 & 2.64 & 11.39 & -2.83 & $-2.16_{-0.06}^{+0.05}$ \\
M0416 & 1843 & 64.02198,\,-24.07786 & 0.4054 & 211 & 1.33 & 10.80 & -3.46 & $-2.49_{-0.22}^{+0.21}$ \\
M0416 & 1906 & 64.02498,\,-24.07208 & 0.3977 & 246 & 1.57 & 11.02 & -2.91 & $-2.06_{-0.06}^{+0.05}$ \\
M0416 & 1920 & 64.02547,\,-24.08509 & 0.3925 & 159 & 2.95 & 10.84 & -2.96 & $-2.27_{-0.14}^{+0.16}$ \\
M0416 & 1975 & 64.02831,\,-24.07228 & 0.3926 & 193 & 1.87 & 10.91 & -3.12 & $-1.64_{-0.12}^{+0.12}$ \\
M0416 & 2003 & 64.02942,\,-24.07902 & 0.4017 & 139 & 3.02 & 10.84 & -2.80 & $-1.87_{-0.11}^{+0.10}$ \\
M0416 & 2008 & 64.02969,\,-24.08344 & 0.3971 & 232 & 6.23 & 11.66 & -2.30 & $-1.88_{-0.03}^{+0.03}$ \\
M0416 & 2151 & 64.03474,\,-24.06981 & 0.4011 & 146 & 3.84 & 10.99 & -2.71 & $-1.82_{-0.29}^{+0.28}$ \\
M0416 & 2176 & 64.03573,\,-24.06967 & 0.3996 & 241 & 5.82 & 11.37 & -3.47 & $-2.31_{-0.17}^{+0.20}$ \\
M0416 & 2201 & 64.03687,\,-24.08066 & 0.4035 & 223 & 3.15 & 11.25 & -2.45 & $-2.03_{-0.03}^{+0.03}$ \\
M0416 & 2239 & 64.03823,\,-24.07176 & 0.3958 & 205 & 2.40 & 11.05 & -3.07 & $-2.28_{-0.07}^{+0.08}$ \\
M0416 & 2246 & 64.03853,\,-24.06208 & 0.4055 & 128 & 3.31 & 10.91 & -2.76 & $-1.98_{-0.10}^{+0.10}$ \\
M0416 & 2270 & 64.03944,\,-24.06932 & 0.4070 & 182 & 2.27 & 10.90 & -2.71 & $-1.82_{-0.03}^{+0.03}$ \\
M0416 & 2317 & 64.04182,\,-24.06282 & 0.3953 & 103 & 1.32 & 10.26 & -3.53 & $-1.54_{-0.37}^{+0.39}$ \\
M0416 & 2333 & 64.04251,\,-24.06923 & 0.3996 & 221 & 2.61 & 11.18 & -2.72 & $-2.20_{-0.02}^{+0.02}$ \\
M0416 & 2335 & 64.04266,\,-24.06514 & 0.3939 & 116 & 3.57 & 10.71 & -2.50 & $-1.75_{-0.17}^{+0.17}$ \\
M0416 & 2336 & 64.04276,\,-24.07303 & 0.4060 & 145 & 2.20 & 10.77 & -2.82 & $-2.11_{-0.07}^{+0.07}$ \\
M0416 & 2387 & 64.04485,\,-24.07352 & 0.4023 & 221 & 5.29 & 11.45 & -2.42 & $-2.21_{-0.01}^{+0.01}$ \\
M0416 & 2395 & 64.04519,\,-24.06213 & 0.4050 & 216 & 2.12 & 11.05 & -3.03 & $-2.21_{-0.03}^{+0.03}$ \\
M0416 & 2408 & 64.04612,\,-24.06395 & 0.3976 & 108 & 2.06 & 10.59 & -3.96 & $-1.97_{-0.25}^{+0.30}$ \\
M0416 & 2412 & 64.04638,\,-24.06708 & 0.3973 & 189 & 4.17 & 11.21 & -2.69 & $-2.19_{-0.03}^{+0.03}$ \\
M1149 & 5 & 177.39110,\,22.40491 & 0.5338 & 259 & 2.36 & 11.32 & -2.87 & $-2.51_{-0.67}^{+0.45}$ \\
M1149 & 44 & 177.40103,\,22.39788 & 0.5410 & 306 & 3.44 & 11.56 & -3.14 & $-2.28_{-0.65}^{+0.61}$ \\
M1149 & 52 & 177.40358,\,22.39638 & 0.5309 & 257 & 2.70 & 11.26 & -2.59 & $-1.88_{-0.33}^{+0.29}$ \\
M1149 & 62 & 177.40646,\,22.38958 & 0.5507 & 290 & 6.47 & 11.83 & -2.46 & $-2.21_{-0.15}^{+0.14}$ \\
M1149 & 68 & 177.40752,\,22.40305 & 0.5400 & 315 & 2.20 & 11.45 & -2.86 & $-2.54_{-0.16}^{+0.19}$ \\
\hline 
     	\end{tabular}
\end{table*}

\begin{figure*}
  \begin{center}
      \includegraphics[width=1.95\columnwidth]{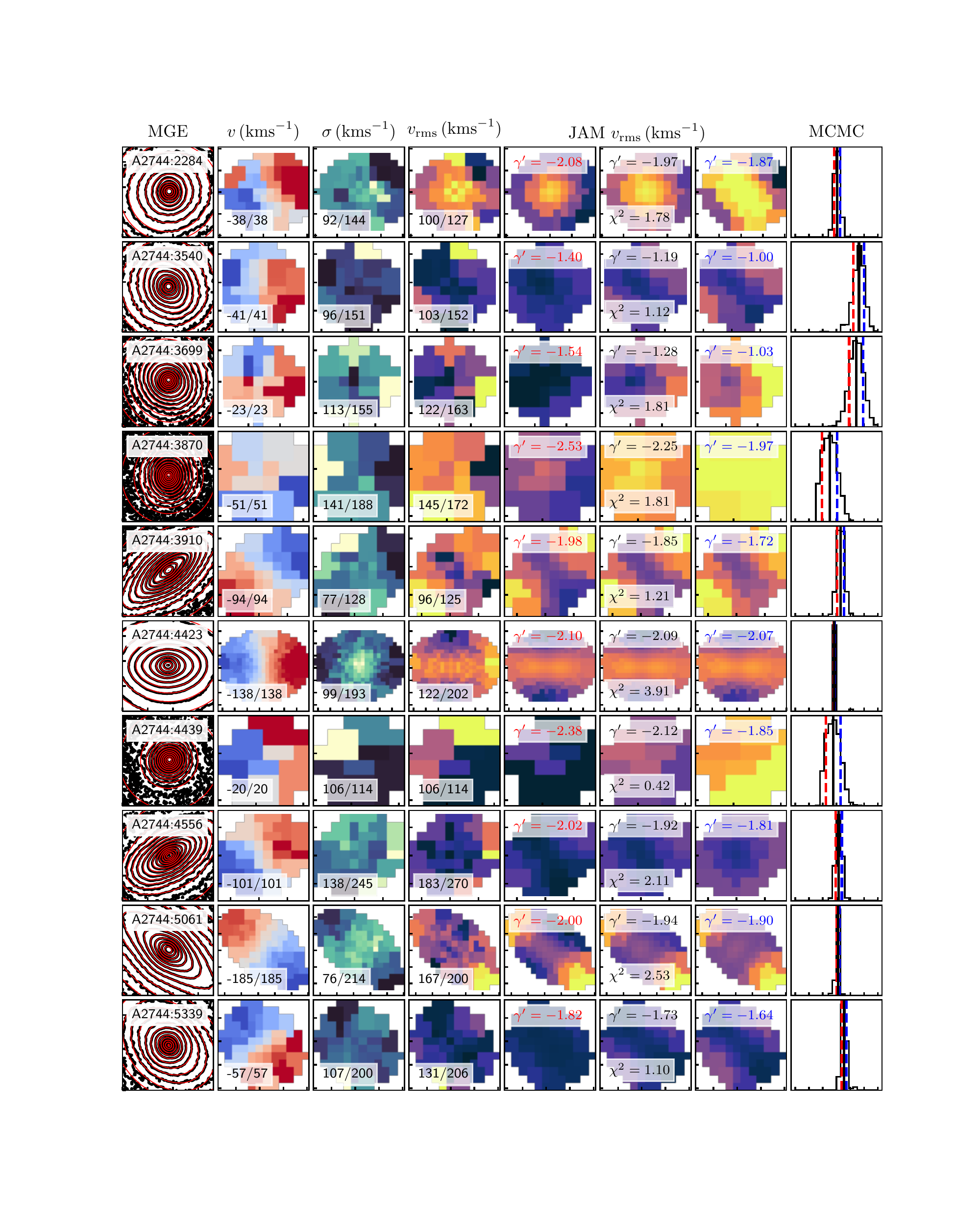}
      \caption{Visual outputs for the 90 galaxies in the sample. Column 1 shows the MGE fit, with the galaxy isophotes in black and the MGE fit in red, in steps of $0.5$ magnitudes. The tick marks indicate 0.5 arcseconds for the MGE and all kinematic fields. The upper corner of this plot gives the cluster and galaxy ID. Column 2 shows the velocity field from pPXF with the colour scale inset. Column 3 shows the velocity dispersion derived from pPXF.  Column 4 shows the observed $v_{\text{rms}}$ field, with the colour scale inset. Columns 5, 6, and 7 show the JAM derived $v_{\text{rms}}$ fields for the 16th, 50th, and 84th percentile inner density slopes respectively.  The fields are created by taking the relevant percentile of \textit{all} parameters; see Figure \ref{longcorner}. These columns share a common colour scale, and the reduced chi-square value is inset on the median field. Column 8 shows the marginalised distribution of the inner density slopes $\gamma'$ from \textsc{emcee}, with the x-axis spanning $-3.5$ to $-0.5$, with tick steps of $0.5$. The red and blue dashed lines indicate the 16th and 84th percentiles respectively, with the median value indicated by a black dashed line.}
      \label{appendixfigure}
    \end{center}
\end{figure*}

\begin{figure*}
  \begin{center}
      \includegraphics[width=2\columnwidth]{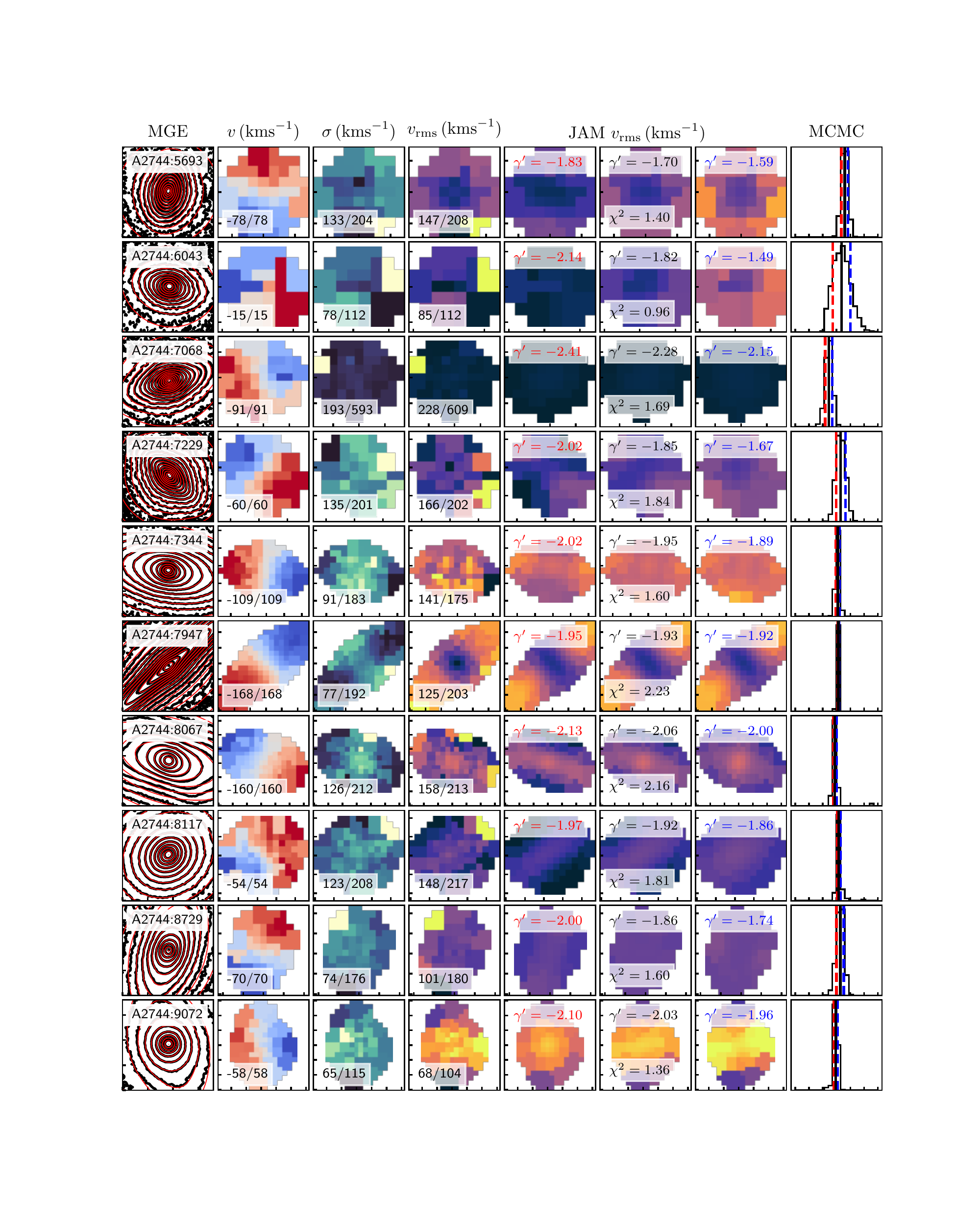}
      \contcaption{}
      \label{appendixfigure1}
    \end{center}
\end{figure*}
\begin{figure*}
  \begin{center}
      \includegraphics[width=2\columnwidth]{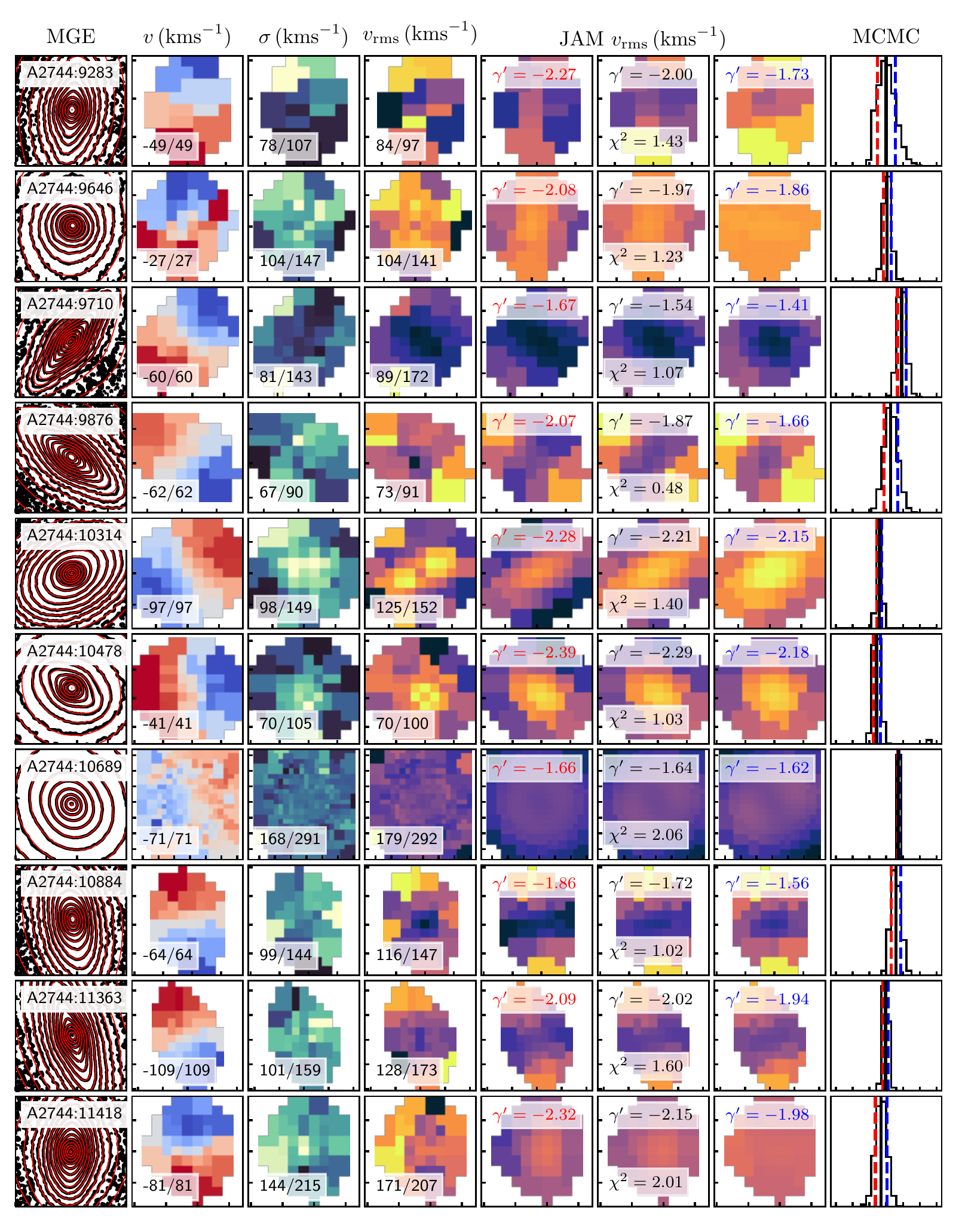}
      \contcaption{}
      \label{appendixfigure2}
    \end{center}
\end{figure*}
\begin{figure*}
  \begin{center}
      \includegraphics[width=2\columnwidth]{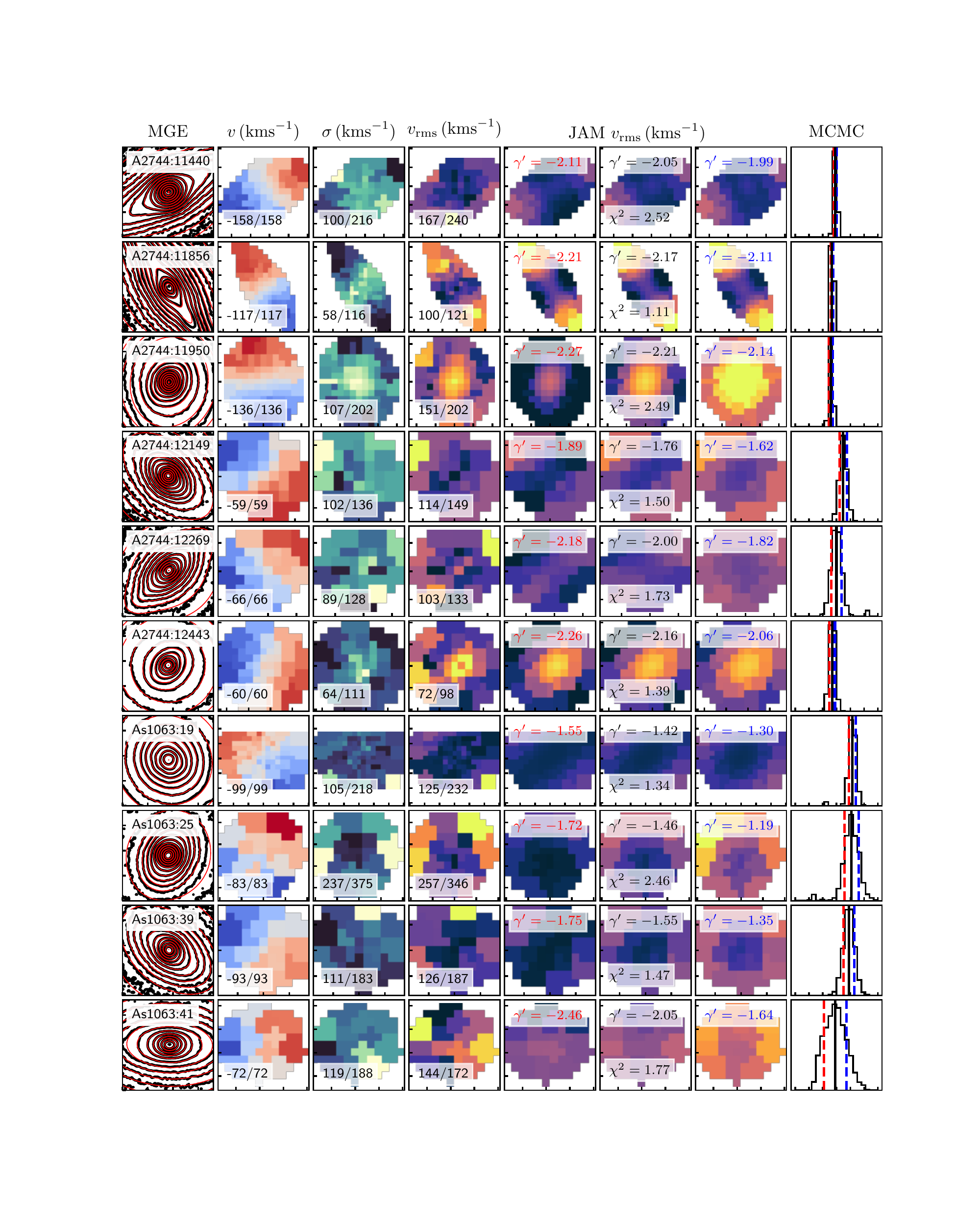}
      \contcaption{}
      \label{appendixfigure3}
    \end{center}
\end{figure*}
\begin{figure*}
  \begin{center}
      \includegraphics[width=2\columnwidth]{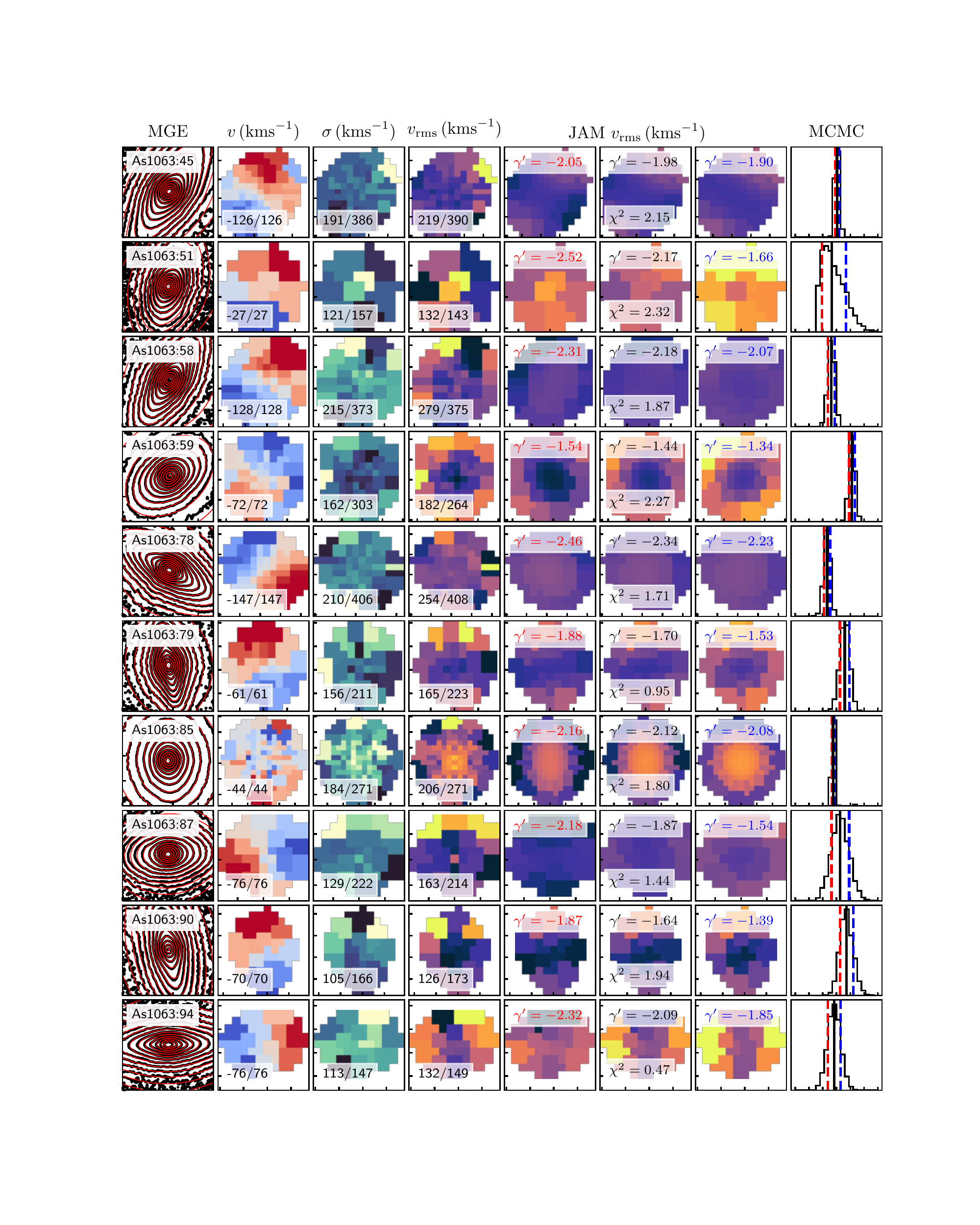}
      \contcaption{}
      \label{appendixfigure4}
    \end{center}
\end{figure*}
\begin{figure*}
  \begin{center}
      \includegraphics[width=2\columnwidth]{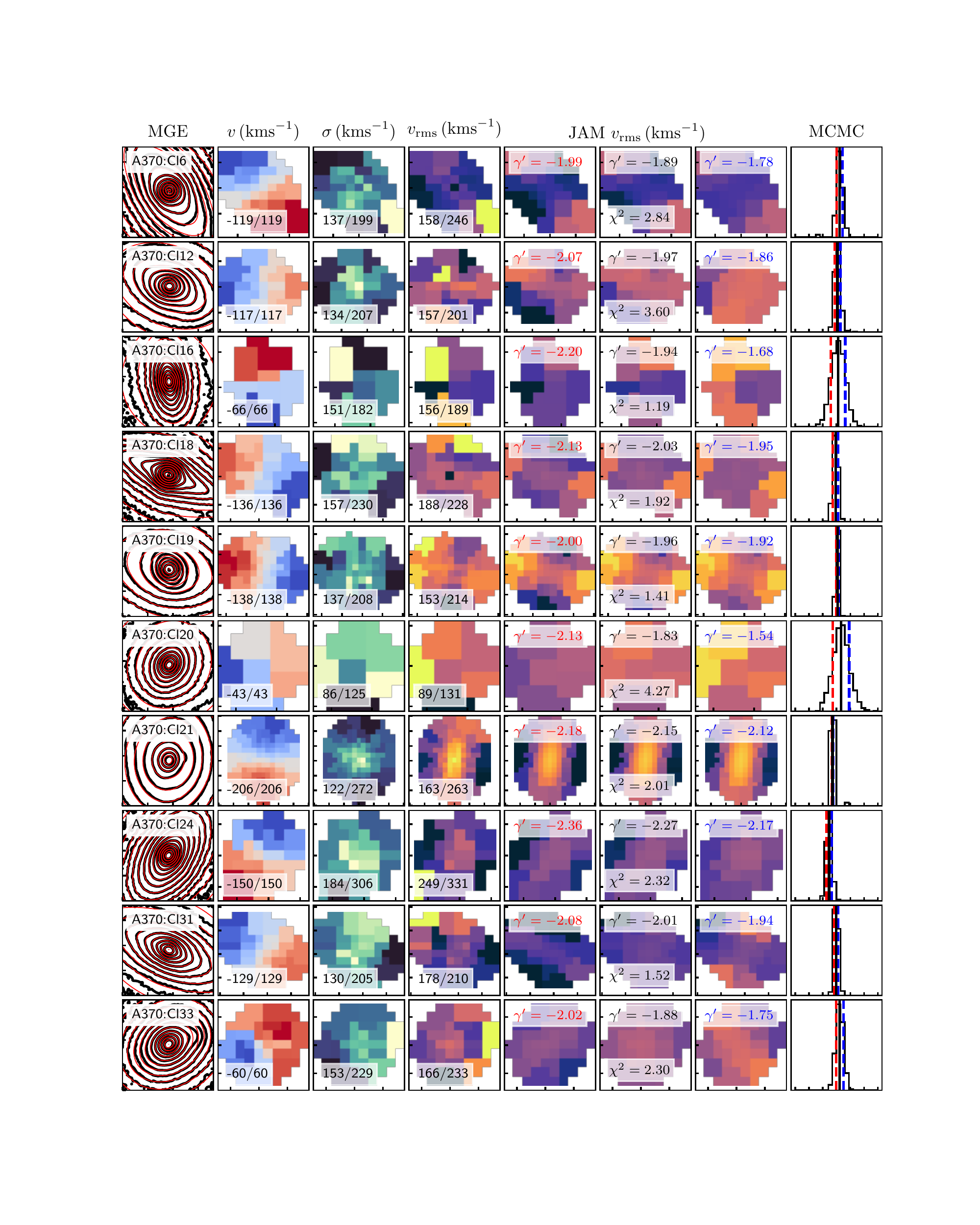}
      \contcaption{}
      \label{appendixfigure5}
    \end{center}
\end{figure*}
\begin{figure*}
  \begin{center}
      \includegraphics[width=2\columnwidth]{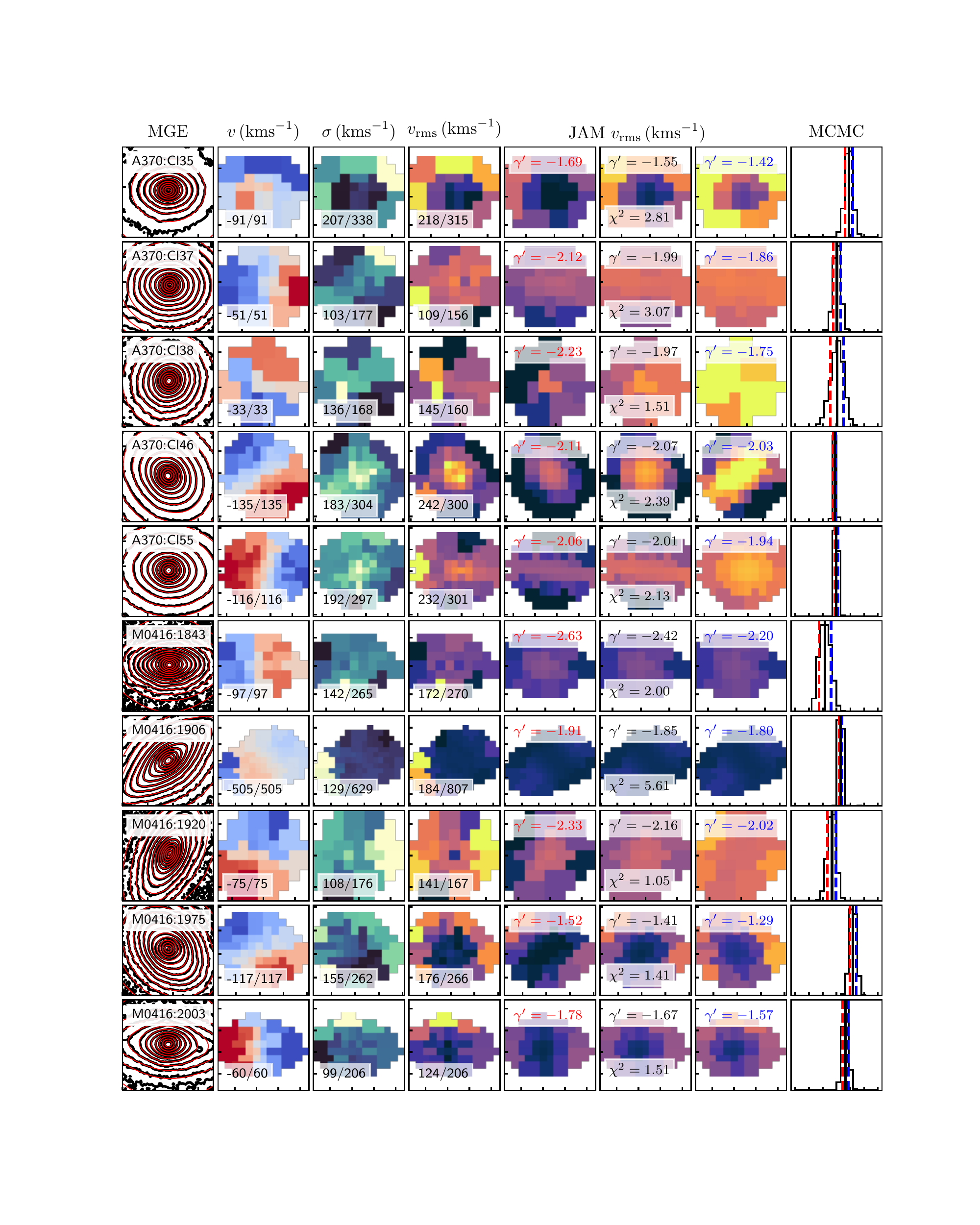}
      \contcaption{}
      \label{appendixfigure6}
    \end{center}
\end{figure*}
\begin{figure*}
  \begin{center}
      \includegraphics[width=2\columnwidth]{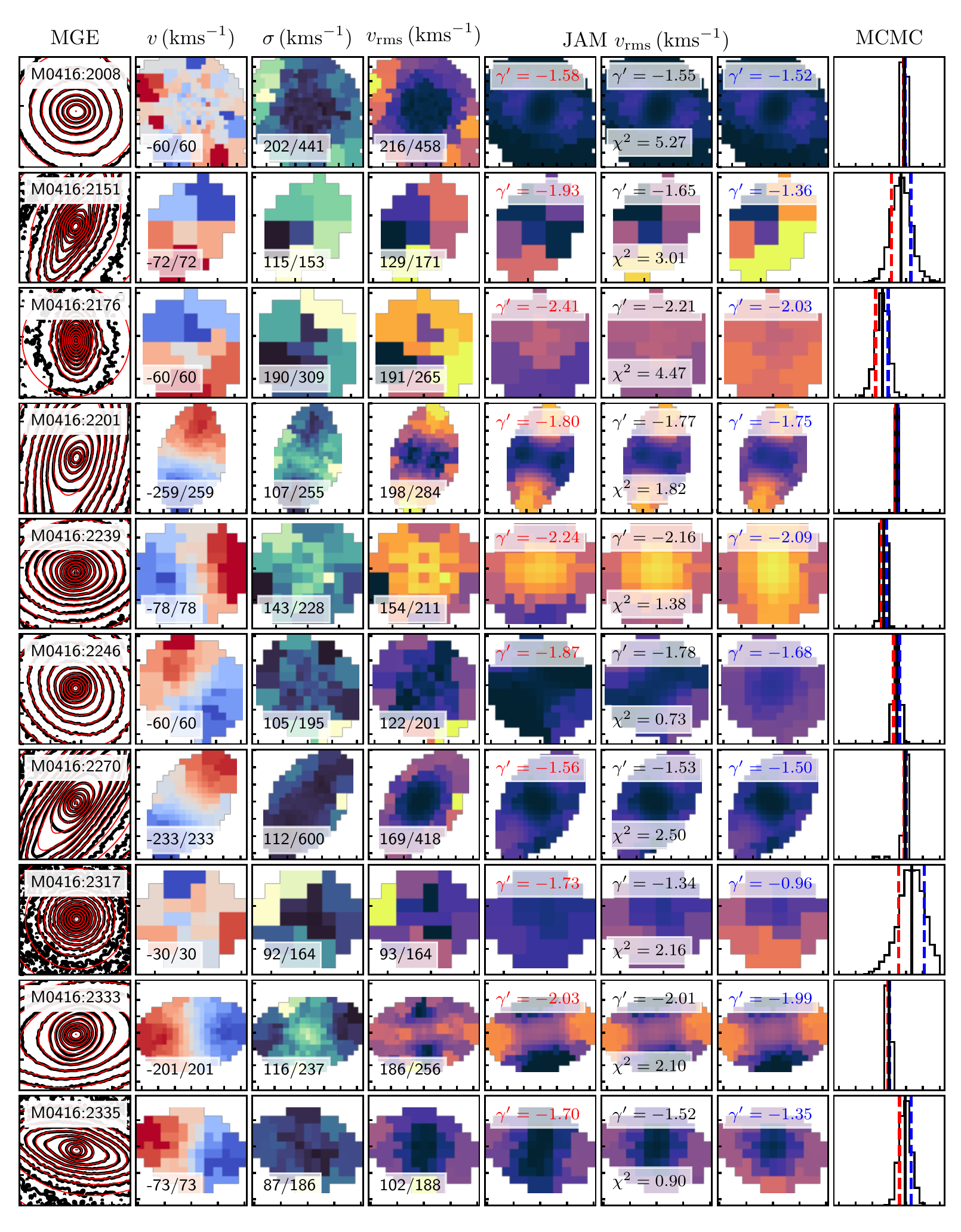}
      \contcaption{}
      \label{appendixfigure7}
    \end{center}
\end{figure*}
\begin{figure*}
  \begin{center}
      \includegraphics[width=2\columnwidth]{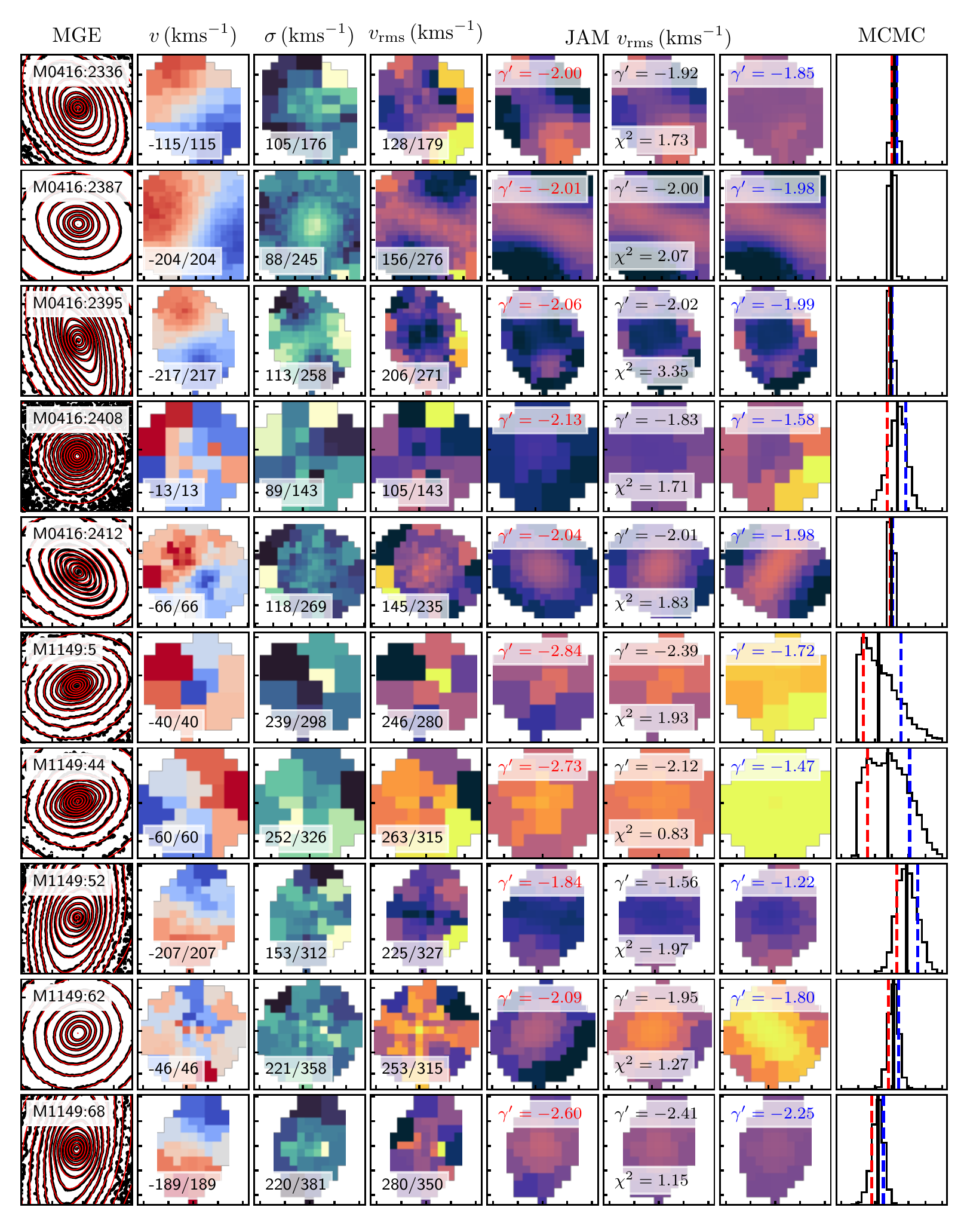}
      \contcaption{}
      \label{appendixfigure8}
    \end{center}
\end{figure*}
\section{simulating the density slope dependence on data quality and PSF}
\label{appendix:simsect}
The salient point of difference between redshift zero studies of total mass density slopes using integral field spectroscopy and this work performed at a higher redshift is the number of spatial elements per kinematic field. This difference can be understood as a result of the cosmological dimming that renders even massive galaxies relatively faint at intermediate redshifts, combined with the reduced angular size of the galaxies at these distances. 
To determine what impact the number of Voronoi bins has on the derived density slope through the \textsc{emcee} process, the MGE of a galaxy with well defined structural parameters was used to create a simulated kinematic field which was subsequently degraded, with the input and output density slopes compared to ascertain the presence of any bias. Further, since many of the galaxies are close to the PSF resolution of the MUSE instrument, the PSF itself becomes a parameter which has the potential to affect the modelled central mass distribution and density slope, and was also investigated through simple Monte-Carlo simulations. The details are as follows. 

The underlying photometry of galaxy 4423 in cluster A2744 was chosen to construct the model surface brightness map, as it has a well fitted MGE and large spatial extent. Additionally, A2744 has three stars in the MUSE field against which the MUSE header FWHM for the PSF can be corroborated, meaning the PSF is particularly well understood for this cluster. The MGE measured on HST pixels was resampled onto a coordinate grid with a MUSE pixel scale of 0.2", with the surface brightness constructed as an `apparent' surface brightness in units of magnitudes per square arcsecond. 

A polynomial fit to empirical surface brightness and SNR measurements was used to give each pixel of the mock galaxy a SNR. The galaxy was then thresholded to a minimum SNR of $2$ per pixel and Voronoi binned to a SNR of 10 per pixel per bin, as was done with the Frontier Fields sample. The best-fit parameters from the \textsc{emcee} models of galaxy 4423 were used to create a total potential model, mapped onto the Voronoi barycentres of the mock galaxy. This total potential model remained fixed for all the simulations. The SNR estimate for each bin was used to create an uncertainty on the $v_{\text{rms}}$ value, randomly drawn, in $\mathrm{kms}^{-1}$.

In essence, this process means that by scaling the surface brightness model, the number of kinematic spatial bins can be changed. The reduction in Voronoi bins happens without changing the underlying total potential model, to mimic data quality becoming worse.

\begin{figure}
  \begin{center}
      \includegraphics[width=\columnwidth]{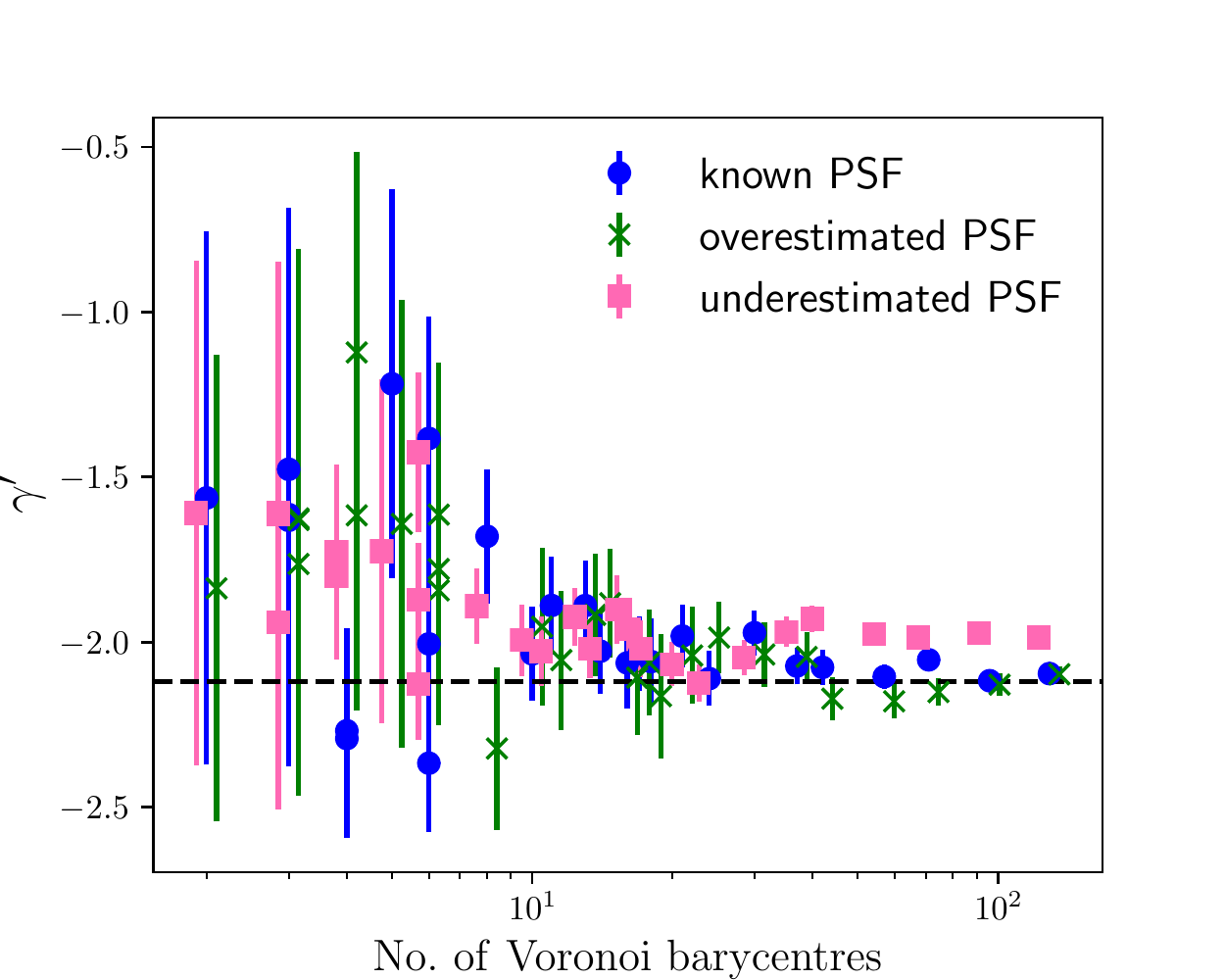}
      \caption{The effect of PSF and degradation of data quality on the recovered density slope is shown. The error bars represent the 16th and 84th percentiles from the obtained \textsc{emcee} distribution. The black dashed line shows the input inner density slope value.}
      \label{simfigure}
    \end{center}
\end{figure}

A range of mock galaxies, in terms of number of Voronoi bins, were created by scaling the underlying photometry. The mock $v_{\text{rms}}$ field was then used to recover the inner density slope using \textsc{emcee} in the same way as for the Frontier Field galaxies, described in Section \ref{jean}. Three sets of models were run: those with no scaling to the PSF, those with the PSF overestimated by $33\%$, and those with the PSF underestimated by $33\%$. The recovered inner density slope from \textsc{emcee} for these simulations are shown in Figure \ref{simfigure}, with the dashed lined indicating the actual inner density slope for all simulations.

With a decreasing number of Voronoi bins, but no uncertainty in the provided PSF, the input density slope was well recovered within the found 1$\sigma$ uncertainties, with 1$\sigma$ uncertainties for $\gamma$ below $\sim 0.2$.  However, when the number of bins dropped to below $\sim 5$, the associated errors encompassed the parameter space with no constraint on the density slope (effectively a uniform distribution). No measurable bias was introduced; while the returned density slopes all became shallower for low bin numbers, this is only because the median value of the parameter space was returned. A threshold of at least 5 bins per kinematic field was used to select the initial galaxy data sample from the MUSE archival data, described in Section \ref{kinematics}. It should be emphasised here that all data cases where the inner density slope was not constrained within the parameter spaces were excluded, and so the bias towards shallower slopes as a result of data quality seen in these simulations does not affect the results of this work.

For the PSF simulations, a 30\% over- and under-estimation was used to test the impact of potential (pessimistic) errors on the PSF used. The results of this are shown in Figure \ref{simfigure}.  The same lack of constraint for the returned density slope was noted at the level of less than five Voronoi bins. For an overestimated PSF, no bias is apparent in the derived density slopes, and the slopes were again well recovered within the $1\sigma$ uncertainties when more than five bins were fitted. However, there is evidence for a mild bias for an underestimated PSF, with estimated uncertainties that do not encompass the input density slope. 

While this suggests that an underestimated PSF leads to a modelled field that allows for shallower central regions, the effect was a shift in slope from $\sim -2.1$ to $\sim - 2$. MGE fits to three stars in the Abell 2744 field in comparison to the MUSE header quoted FWHM indicate the error on the FWHM is in fact much smaller than the 33\% under or overestimation used in the simulations, on the order of 10\%, with a variation across the field of less than 5\%. Based on the results of the simulations, it is not expected that the PSF values used in this work  on the observed sample artificially shifted the derived density slopes by any significant amount, nor did the use of few spatial elements introduce a bias. 

% Don't change these lines
\bsp	% typesetting comment
\label{lastpage}
\end{document}